\documentclass{aa} 
\usepackage{graphicx}
\usepackage{soul} 

\usepackage{natbib}
\bibpunct{(}{)}{;}{a}{}{,} 

\newcommand{\angstrom}{\mbox{\normalfont\AA}}
\newcommand{\kms}{km\,s$^{-1}$}

\begin{document} 

   \title{Launching the asymmetric bipolar jet of DO Tau} 

   \subtitle{} 

   \author{J. Erkal 
          \inst{1}
          \and C. Dougados
         \inst{3}
         \and D. Coffey 
         \inst{1,2}
          \and S. Cabrit
          \inst{4,3}
          \and F. Bacciotti
          \inst{5}
          \and R. Garcia-Lopez
          \inst{1}
          \and D. Fedele
          \inst{5,6}
          \and A. Chrysostomou
          \inst{7}
            }
   \institute{School of Physics, University College Dublin, Dublin 4, Ireland \\
              \email{jessica.erkal@ucdconnect.ie}         
             \and School of Cosmic Physics, The Dublin Institute for Advanced Studies, Dublin 2, Ireland 
             \and Univ. Grenoble Alpes, CNRS, IPAG, 38000 Grenoble, France
             \and Observatoire de Paris, PSL University, Sorbonne Université, CNRS, LERMA, F-75014, Paris, France
             \and INAF, Osservatorio Astrofisico di Arcetri, Largo E. Fermi 5, 50125, Firenze, Italy
             \and INAF, Osservatorio Astrofisico di Torino, Via Osservatorio 20, 10025, Pino Torinese, Italy
             \and SKA Organisation, Jodrell Bank Observatory, Lower Withington, Macclesfield, Cheshire, SK11 9FT, UK 
        }
   \date{}

\abstract{The role of bipolar jets in the formation of stars, and in particular how they are launched, is still not well understood. }
{We probe the protostellar jet launching mechanism, via high resolution observations of the near-IR [Fe {\sc ii}]$\lambda$1.53,1.64~$\mu$m emission lines.}
{We consider the case of the bipolar jet from Classical T Tauri star, DO Tau, and investigate the jet morphology and kinematics close to the star (within 140 au), using AO-assisted IFU observations from GEMINI/NIFS. 
}
{We find that the brighter, blue-shifted jet is collimated very quickly after it is launched. This early collimation requires the presence of magnetic fields. 
We confirm velocity asymmetries between the two lobes of the bipolar jet, and also confirm no time variability in the asymmetry over a 20 year interval. This sustained asymmetry is in accordance with recent simulations of magnetised disk-winds. 
We examine the data for signatures of jet rotation. We report an upper limit on differences in radial velocity of 6.3 and 8.7~km~s$^{-1}$ for the blue and red-shifted jets, respectively. Interpreting this as an upper limit on jet rotation implies that any steady, axisymmetric magneto-centrifugal model of jet launching is constrained to a launch radius in the disk-plane of r$_0$ $<$ 0.5 and 0.3~au for the blue and red-shifted jets, respectively. This supports an X-wind or narrow disk-wind model. However, the result pertains only to the observed high velocity [\ion{Fe}{II}] emission, and does not rule out a wider flow launched from a wider radius. We report detection of small amplitude jet axis wiggling in both lobes. We rule out orbital motion of the jet source as the cause. Precession can better account for the observations but requires double the precession angle, and a different phase for the counter-jet. Such non-solid body precession could arise from an inclined massive Jupiter companion, or a warping instability induced by launching a magnetic disk-wind.}
{Overall, our observations are consistent with an origin of the DO Tau jets from the inner regions of the disk.} 


\maketitle

\section{Introduction}


Jets and outflows from newly forming stars are a commonly observed phenomenon, but whether they play a critical role in the star formation process is still an open question. Principally, they are invoked to explain the removal of angular momentum from the newly forming star-disk system, allowing the star to spin-down and the disk to accrete. Furthermore, if jets do indeed play this critical role, identifying the launching mechanism may have implications for the disk physics in the region where planets are formed (see reviews by \citet{Baruteau2014} and \citet{Turner2014}). 


Early magnetohydrodynamic (MHD) models of bipolar jets, based on the work of \citet{BlandfordPayne1982}, reveal a link between jet launching and the magnetic fields threading the accretion disk \citep[for e.g.][]{Shu2000,Ferreira1997}. Furthermore, mass outflow and angular momentum removal can occur along magnetic field lines within the inner regions of the disk (\citealp{UchidaShibata1985}). \citet{CasseKeppens2002} present the first global simulations of jet launching from a resistive disk, finding that collimation occurs as the jet propagates outwards from the star at a nearly constant rate, and that the jet removes a significant portion of the energy released by accretion \citep{CasseKeppens2004}. While the theory is broadly supported by numerical simulations, observational support has been tricky to obtain due to the small spatial scales involved (i.e. tens of au, equivalent to sub-arcseconds for the nearest star-forming regions).




Recent decades have seen ever higher spatial and spectral resolution observations provide much needed constraints on proposed models by probing the initial jet channel i.e. within 100 au from the star (see review by \citet{Frank2014}). 
For example, the jet collimation zone ($<$100 au from the star) reveals that jets from both T Tauri (Class II) stars and younger sources (Class 0 and I) are already collimated by the time they propagate to z=50 au (\citealp{Dougados2000};  \citealp{Cabrit2007}). These observed similarity in jet widths, independent of evolutionary stage, suggest magnetic collimation processes which in turn implies a magnetohydrodynamic (MHD) jet launch model (e.g. \citealp{Cabrit2007}). 
Indeed, most recently, cylindrical collimation as early as z=30 au has been observed in the DG Tau jet (\citealp{Bacciotti2002}; \citealp{Maurri2014}). 


In addition, the initial jet channel has been observed to trace out a systematic wiggling pattern in a number of cases (see for example HH~30 \citep{Anglada2007}, HH~24 \citep{Terquem1999}, ZCMa \citep{Antoniucci2016}, RY Tau \citep{Garufi2019}). 
The fact that this wiggle pattern is observed close to the star suggests that, rather than being an effect of interactions with the environment, it is either produced by orbital motion of the jet source in the binary system, see e.g. the study of the HH~30 bipolar jet wiggling by  \citep{Anglada2007} or that the cause of the jet wiggling is intrinsic to the launching mechanism.


Furthermore, observations have identified asymmetries between the two lobes of the bipolar jet in terms of velocity, density and opening angle as well as time-variability in velocity. Examples include the bipolar jets from HH~30, RW Aur, DO Tau, DP Tau, DG Tau and DG Tau B (\citealp{Hirth1994}; \citealp{Woitas2002} \citealp{Melnikov2009}; \citealp{Hartigan2009}; \citealp{Agra-Amboage2011};  \citealp{Podio2011}). The question is whether bipolar asymmetries originate in the launching mechanism itself, or as a result of interaction with an asymmetric environment. If such asymmetries could be observed soon after launching (i.e. $<$ 100 au from the star), this would render the latter less likely. 


Numerical simulations have risen to the challenge of reproducing observed jet wiggling and other asymmetries. Studies have investigated the launching of intrinsically asymmetric high velocity jets from stellar magnetospheres \citep[for e.g.][]{Lovelace2010,Lii2014,Dyda2015} and magnetised disks \citep{Fendt2013}. It has been shown that the presence of a companion star or interactions with the interstellar magnetic field can cause the jet axis to curve and bend \citep{FendtZinnecker1998}. 
However, many simulations consider the larger scales. Few focus on the initial jet channel which is more likely to better constrain the launching mechanism. Examples include recent 3D MHD simulations of jet precession caused by a warped disk in a binary system (\citealp{SheikhnezamiFendt2015}; 2018). These latter studies find that the differential tidal effects of a close companion could generate asymmetries between jet and counter-jet collimation and axis wiggling.



Intriguingly, observational support for magneto-centrifugal jet launch models was also pursued through investigations of kinematic signatures interpreted as jet rotation (and hence evidencing angular momentum transport). This wave of studies was initiated by HST observations of Classical T Tauri jets in optical and near-UV atomic tracers (\citealp{Bacciotti2002}; \citealp{Coffey2004}, 2007, 2012), and was soon followed by similar studies from the ground in near-IR atomic lines (\citealp{Chrysostomou2008}; \citealp{White2014}, \citealp{Coffey2011}, \citealp{Coffey2015}). However, debate continues over the interpretation of observed kinematic signatures which may have other causes, among them asymmetric shocks and/or jet wiggling (\citealp{Soker2005}; \citealp{Cerqueira2006}; \citealp{Fendt2011}; \citealp{Staff2012}). Most recently, the high resolution of ALMA helped lend excitement to claims of rotation signatures in molecular tracers of high velocity jets (\citealp{Lee2017}). This assumes, of course, that the molecules can actually survive the MHD launching process \citep{Panoglou2012}. Then, the main concern when using molecular versus atomic tracers is that molecular emission may be contaminated or dominated by emission from the cavity walls and entrained ambient material, rather than the flow itself (e.g. \citealp{Hirota2017}). Thus, a combined approach is required, whereby the atomic and molecular tracers are observed at several wavelengths to give a complete picture \citep{Coffey2017}. 



Clearly, observations close to the jet base are critical to probe intrinsic launching signatures and minimise ambient contamination. Such jet studies within 100 au of the star are observationally demanding, requiring sub-arcsecond resolution often in combination with a long slit or integral field unit. So inevitably they occur on a case-by-case basis, and for the brightest targets. In this vein, here we present the case of Classical T Tauri star, DO Tau, using adaptive-optics assisted spectro-imaging observations of near-infrared atomic tracers, close to the base of the flow.

DO Tau, lies in the Taurus cloud at a distance, $d$, of 139$\pm$7 pc \citep{GaiaCollaboration2018}. We adopt the stellar properties listed in \citet{Long2019}: stellar mass $M_{\star} \sim$ 0.59~$M_{\sun}$, stellar age $t_{\star}$ $\sim$ 5.9 $\rm Myr$ and extinction $A_v \sim 0.75$. Spectral type is given as M0.3 \citep{Herczeg2014}. From modelling of the UV excess flux, \citet{Gullbring1998} derives a disk accretion rate, $M_{acc}$ $\ge$ 1.44 $\times$ 10$^{-7}$ $M_{\sun}yr^{-1}$. Similarly, \citet{Eisner2014} derives from the observed Br$_\gamma$ flux an accretion luminosity $L_{acc}$ $\sim$ 1.5 $L_{\sun}$, yielding a mass accretion rate $M_{acc}$ $=$ 2.1 $\times$ 10$^{-7}$ $M_{\sun}yr^{-1}$ with the adopted stellar parameters. However, recently, \citet{Simon2016} derived from optical veiling measurements an order of magnitude lower L$_{acc}$ ($\sim$ 0.1 $L_{\sun}$), suggesting possible strong accretion variability in this source. 

The disk of DO Tau was recently mapped with ALMA in the continuum at high angular resolution by \citet{Long2019}. The dust continuum emission was modelled to reveal a compact, smooth disk with effective radius, encompassing 95\% of encircled flux, of $r_{\rm eff,95} =$ 0$\farcs$263 (37 au), inclination angle of $i_{\rm disk}$ $\sim$ 27.6$^{\circ}$ $\pm$ 0.3 to the line-of-sight, and position angle, PA$_{\rm disk}$ $\sim$ 170$^\circ$ $\pm$ 0.9. The quoted uncertainties on the disk parameters are formal uncertainties in the fitting procedure. True uncertainties are probably larger especially in the case of a face-on disk. Recently, \citet{FernandezLopez2020} derived an inclination angle of 19$^{\circ}$ from ALMA data. We therefore adopt an inclination of 23.5$^{\circ}$ (the average of \citet{Long2019} and \citet{FernandezLopez2020}) and a more conservative estimate of 5$^{\circ}$ for the disk inclination uncertainty. 

The jet from DO Tau was originally revealed through optical forbidden emission lines \citep{Edwards1987}. A radial velocity asymmetry in the bipolar jet was identified of -92 and +210 km~s$^{-1}$ at 1$\farcs$5 along the blue- and red-shifted jets respectively in [\ion{S}{II}]
\citep{Hirth1997}. More recent observations, which spectrally resolve high and low velocity components of the blue-shifted jet, show no velocity variation over time for a given tracer (\citealp{Simon2016}; \citealp{Giannini2019}), and reveal plasma conditions in the inner 100 au from the disk surface, including a typical electron temperature ($T_e \sim$ 8000\,K), a lower than usual ionisation fraction ($x_e <<$ 0.1), and a hydrogen number density in the low and high velocity components $n_H \sim 10^{6-7} {\rm cm}^{-3}$ \citep{Giannini2019}. Most recently, \citet{FernandezLopez2020} report a ringed pole-on outflow based on CO data from ALMA. 



On parsec scales, DO Tau is associated with HH objects HH~831A, HH~831B and HH~832 \citep{McGroarty2004}, all of which are located to the northeast as far as 11$'$ i.e. $\sim$~90~pc. With position angles of $\sim$74-78$\degr$, they are well-aligned with the red-shifted jet (PA$_{jet}\sim$70$\degr$, \citealp{Hirth1997}). Also coinciding with this position angle is an extended reflection nebulosity to the north-east of DO Tau \citep{Magakian2003}. Observed near-infrared emission, extending to the northeast as far as 1$\farcs$7 with an inner boundary at 1$\farcs$1, was interpreted as scattering from a cleared cavity \citep{Itoh2008}. Observed far-infrared emission extending in the same direction \citep{Howard2013}, forms an arc which connects DO Tau to neighbouring Classical T Tauri star, HV Tau, to the east at $\sim$7$'$ ($\sim$1\,kpc). When interpreted as a 'bridge' of dust emission, it prompted the suggestion of a disk-disk encounter \citep{Winter2018}. 

Here, we present our study of the DO Tau jet and disk system at high resolution. In section \ref{sec:obs}, we outline the observations and archival data presented. In section \ref{sec:data}, we describe the data reduction and analysis. In section \ref{sec:results}, we set out our main findings and in section \ref{sec:discussion} we discuss the implications in the context of relevant modelling. Section \ref{sec:conclusions} summarises our conclusions. 



\section{Observations and data reduction}
\label{sec:obs}

\subsection{NIFS observations}

The bipolar jet of T Tauri star DO Tau was observed in the H-band using the Gemini-North NIFS instrument, with Altair adaptive optics (AO) correction (Program ID: GN-2009B-Q-43), on 2009 Nov 11, Dec 17, 23 and 24. 

The star was offset from the center of the NIFS field-of-view (3$\arcsec$ $\times$ 3$\arcsec$) in order to include as much of the bright blue-shifted jet as possible while also observing the stellar PSF. This was to facilitate a PSF deconvolution of the jet image, to maximise the spatial resolution of the jet. 
Fortunately, the dimmer red-shifted jet also appeared within the field-of-view. 

NIFS spatial sampling is 0$\farcs$043 $\times$ 0$\farcs$103. The direction with the smallest spatial sampling (detector x-axis) was aligned perpendicular to the jet position angle, in order to achieve the highest spatial sampling in the direction transverse to the jet axis. This corresponded to an instrument position angle, PA$_{\rm{IFU}}$=160$^{\circ}$. Observations were also conducted with an anti-parallel position (i.e. PA$_{\rm{IFU}}$=340$^{\circ}$), to facilitate identification of any contamination by PSF artefacts. The use of AO correction yielded angular resolution of FWHM=0$\farcs$15-0$\farcs$175, determined by measuring the FWHM of the continuum. 
The NIFS H-band covers 1.49-1.80 $\mu$m, which includes the jet emission lines of [\ion{Fe}{II}] 1.53, 1.60 and 1.64 $\mu$m. NIFS spectral sampling is 1.6~\text{\AA}, giving a two-pixel velocity resolution of 56 \kms. 
The observations were conducted using the standard ABA nodding technique. A total of 96 short exposures were obtained, reaching an on-target total exposure time of 3.2 hours. 

\subsection{ALMA observations}

Archival ALMA observations were retrieved in order to study the disk rotation sense. We use in this paper observations of the DO~Tau disk in the C$^{18}{\rm O}({\rm J}=2-1)$ line at 219.560~GHz from ALMA program 2016.1.00627.S (P.I. K. Oberg) presented in \citet{Bergner2019}. Observational and data reduction details are given in \citet{Huang2017}. The resulting beam is $0.72^{\prime\prime} \times  0.46^{\prime\prime}$ at PA=30$^{\circ}$ and the sensitivity (1 $\sigma$ rms) is 5~mJy/beam per 0.2~\kms channel. 

\subsection{NIFS data reduction}
\label{sec:data}

The standard GEMINI calibration pipeline was used to reduce the NIFS data. Each night was processed individually. 
Due to a deterioration in observing conditions during one of the four nights, there are fewer exposures for the anti-parallel slit (2009 Dec 23, PA$_{\rm{IFU}}$=340$^{\circ}$). The exposures for this one night were not included in further analysis. A second night (2009 Nov 11, PA$_{\rm{IFU}}$=160$^\circ$) was also excluded due to a time interval of several weeks between it and the other observations. In this data, a small change in jet velocity of 5-10 km~s$^{-1}$ was measured. Such small variability in jet velocity is expected, especially when there is a time interval of several weeks between observations. However, since we are investigating small changes in velocity on this order, it was deemed prudent to discard these exposures.  

For the two remaining nights, 2D Gaussian fits to the individual continuum images were used to exclude exposures with poorer seeing, in an attempt to maintain higher spatial resolution. Although parallel and anti-parallel slit positions were requested in order to identify artefacts, the challenge of achieving sufficient signal-to-noise in the jet borders meant we were compelled to combine these two slit orientations. Any artefacts would be cancelled out in this way, with a resulting improvement in the quality of the data. 

Thus, since we used only two of four datacubes (2009 Dec 17 (PA=340$^\circ$) and Dec 24 (PA=160$^\circ$)), we achieve a total on-target exposure time of about one hour, and a PSF of 0$\farcs$15. 

To form the datacubes, co-adding the individual exposures first required aligning the exposures via a 2D Gaussian fit of the stellar PSF. Then, co-adding was conducted via sub-pixel shifts using the IRAF tool {\em imshift}. The data were flux calibrated using the telluric standards. 

A dedicated continuum subtraction procedure was employed to optimize subtraction of the bright stellar contribution and better isolate the faint jet line emission in its immediate vicinity. We followed the method outlined in \citet{Agra-Amboage2011}. This method allows us to correct both from the photospheric spectrum and telluric absorption features close to the central source. However, it also removes any unresolved [\ion{Fe}{II}] emission close to the star, and so we cannot study [\ion{Fe}{II}] closer than 0$\farcs$1 to the source. The datacubes were then rotated to align the jet axis with the horizontal, to facilitate eventual analysis using 1D Gaussian fitting. 

The next steps in the data reduction saw the procedure branch in two, with one approach deemed more appropriate for analysis of the morphology and the other for the kinematics. To facilitate analysis of the jet morphology, we first combined the datacubes for the parallel and anti-parallel slit positions. We then spectrally binned the [\ion{Fe}{II}] 1.64 $\mu$m line, integrated over v$_{\rm LSR}$=[-122-228]~$\rm km~s^{-1}$, to create 2D images of the approaching and receding jets, respectively. We then further improved angular resolution via deconvolution using the IRAF \textit{Lucy} routine. The PSF was constructed by summing the continuum emission, near the [\ion{Fe}{II}] 1.64 $\mu$m line, over a number of spectral channels similar to the width of the [\ion{Fe}{II}] 1.64 $\mu$m line (5 channels). The deconvolution improved resolution further to $\le$~0$\farcs$06 (FWHM). For both the jet and counter-jet images, we conducted 1D Gaussian fitting across the jet intensity profile, yielding jet axis position and jet width as a function of projected distance from the source. 

Meanwhile, for analysis of the jet kinematics, PSF deconvolution was avoided for fear of contaminating the kinematic signature. However, critically, a correction was required for a possible uneven slit illumination effect. Non-uniform illumination within an IFU slitlet introduces a wavelength shift in the spectrum at that position which can be up to a few km~s$^{-1}$, \citep{Agra-Amboage2014}. The uneven slit correction was conducted on datacubes for both the parallel and anti-parallel IFU positions, PA$_{{\rm IFU}}$, before combining them. We found this to be the optimum procedure to remove all possible instrumental effects, and fundamental to a jet rotation study. This is because the differences in radial velocity under investigation are on the same order as those expected from uneven illumination effects, i.e. a few km~s$^{-1}$. The uneven slit illumination effect was modelled in each IFU spaxel by estimating the displacement of the centroid of the brightness distribution within the spaxel with respect to the slitlet center. From this a 2D velocity map correction was estimated and subtracted from the observed 2D centroid velocity maps. 
As a double-check, the uneven slit 2D correction map was also computed
from a photospheric absorption line near the [\ion{Fe}{II}] 1.64 $\mu$m emission, by estimating at each spaxel position the velocity shift of the centroid of the line with respect to the spectrum located at the continuum centroid position (i.e. the defined star center). 
 The required correction was found to be typically of a few km~s$^{-1}$ and up to $\pm$ 8~km~s$^{-1}$ at 0$\farcs$1 from the star (see Appendix, figure \ref{fig:uneven_slit}).
 

The datacubes for the parallel and anti-parallel slit positions were then combined, and Gaussian profile fitting was conducted at each spatial position to retrieve the jet peak centroid velocity map in [\ion{Fe}{II}] 1.64 $\mu$m. Lastly, all velocities were corrected to the LSR velocity frame, using for DO Tau the systemic LSR velocity of +6.47 $\pm$ 0.14 km~s$^{-1}$ \citep{Guilloteau2016}. 

The differences in radial velocity under investigation are expected to be on the same order as the residuals from wavelength calibration. Although the absolute velocity calibration accuracy is not critical for this study, the relative accuracy within the IFU spaxels and over a wavelength range close to the line of interest is crucial. A detailed analysis of telluric sky emission lines close to the [Fe {\sc ii}]1.64 $\mu$m line is described in a similar study for RY Tau \citep{Coffey2015}, which was observed during the same period and with the same instrument configuration. From this study, we estimate an internal relative calibration 1$\sigma$ error of $\sim$ 1.6 $\rm{km~s^{-1}}$ \citep{Coffey2015} over the IFU field of view.  We also estimate a 1$\sigma$ error on the global absolute velocity calibration scale of 10~\kms. 
When performing Gaussian fitting to derive the line velocity centroids, errors in centroid velocity are a combination of the error in velocity calibration and that of Gaussian fitting. The former remains constant, while the latter is signal-to-noise dependent. Fitting errors were calculated in accordance with the equation for the accuracy of a Gaussian centroid fit as applied to cases of high signal-to-noise such that photon noise dominates, and assuming that the profile is well-sampled \citep{Whelan2008}: 

\begin{equation}
    \centering
    \sigma_{centroid} = \frac{\text{FWHM}}{2\sqrt{2~\text{ln}~2}~\text{SNR}}
\end{equation}

where FWHM represents the width of the fitted profile, (here our spectral resolution of 56 \kms), and SNR is the signal-to-noise ratio at the Gaussian peak. For our observations, we thus derive typical centroid velocity 3$\sigma$ accuracy of 2-5 \kms, depending on the signal to noise. Combining this in quadrature with the internal relative calibration error gives the error bars in figure \ref{fig:velocity_trans}. 

\section{Results}
\label{sec:results}

\subsection{Jet Morphology} 
\label{sec:results_morph}

\begin{figure*}
    \centering
    \includegraphics[width=0.9\textwidth]{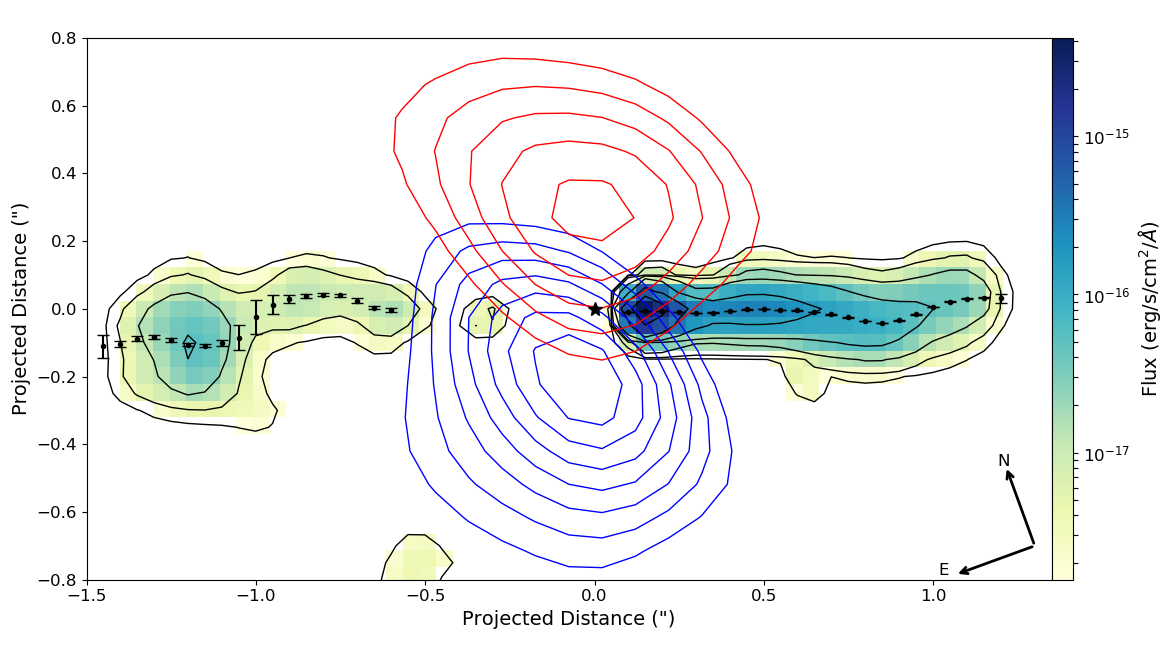}
    \caption{Deconvolved image of the DO Tau receding (left) and approaching (right) jet in \ion{Fe}{II} 1.64 $\mu$m line, integrated over $\text{v}_{\text{LSR}}=[-122,228]~\text{km~s}^{-1}$. The contour floor is $3 \sigma$ increasing in log intervals of 0.5 (where $1 \sigma$ is $1.43~\times~10^{-18}~\text{erg~s}^{-1}~\text{cm}^2~\text{\AA}$. The derived jet axis position is overlaid in black dots with associated error bars. Rotating to the horizontal yields a blue-shifted jet position angle, PA$_{\text{jet}}$ = 260$\degr$. Blue and red contours trace C$^{18}$O emission integrated over $\text{v}_{\text{LSR}}=[4.21,5.01]~\text{km~s}^{-1}$ and $\text{v}_{\text{LSR}}=[7.01,7.61]~\text{km~s}^{-1}$ respectively. The contour floor is $3~\sigma$ (13 mJy/beam), increasing in linear intervals of $1~\sigma$ (4.5 mJy/beam).} 
    \label{fig:morphology}
\end{figure*}

We present the first high spatial resolution images of the DO Tau bipolar jet close to the star. Figure \ref{fig:morphology} shows a deconvolved intensity maps of the red- and blue-shifted jets in [\ion{Fe}{II}] 1.64 $\mu$m~emission. (Recall the map is deconvolved as described in section \ref{sec:data}.) The blue-shifted jet is brighter (right side of figure \ref{fig:morphology}). Overlaid are intensity contours of C$^{18}$O molecular emission, from ALMA archival data, which traces the gaseous disk. 

Rotating the jet image to the horizontal provided a revised estimate of the blue-shifted jet PA = 260$\degr \pm$ 3$\degr$ (compared to 250$\degr \pm$ 10$\degr$ \citep{Hirth1997}). This is in very good agreement with the recent determination of \citet{FernandezLopez2020} from HST imaging and with the large scale disk $PA=170 \pm 0.9^{\circ}$ derived by \citet{Long2019} from ALMA high-angular resolution continuum observations. 
There may be a small misalignment of the red-shifted jet PA by $\simeq$ 1$\degr$, but changes in the jet trajectory makes it difficult to claim with certainty. 

The blue-shifted jet is detected as close as 0$\farcs$1 from the star and extends to 1$\farcs$2. The dimmer counter-jet does not appear until 0$\farcs$5, aside from a small emission at $\sim$ 0$\farcs$35, and then similarly extends to 1$\farcs$5. Given the disk inclination angle to the line-of-sight of 28$^{\circ}$ \citep{Long2019}, it seems that the base of the red-shifted jet may be obscured by a disk of projected radius 0$\farcs$5 (70 au), or deprojected radius of 0$\farcs$57 (79 au). \citet{Long2019} derive a dusty disk radius of 0$\farcs$263 (which converts to 39 au), which is very compact with respect to other disks in their survey, and too compact to obscure the red-shifted jet out to 0$\farcs$5. Similarly, \citet{FernandezLopez2020} report a dusty disk radius of 47 au (which converts to 0$\farcs$33). 
We note, however, that continuum disk sizes in ALMA observations only probe millimetre-sized grains, which appear to undergo significant inward migration relative to the gas. On the other hand, near-infrared extinction is dominated by small sub-micron grains expected to remain well-coupled to the gas. Referring to Figure \ref{fig:morphology} of  \citet{FernandezLopez2020}, we see that the Keplerian curve of the $^{12}$CO emission is traced roughly out to 1$\arcsec$ (i.e. 150 au). This shows that the gas disk, and therefore the associated small grains, are extended enough to explain occultation of the base of the red-shifted jet in our data.


The jet image highlights morphological asymmetries between the two jet lobes: the blue-shifted jet appears to maintain the same width as it propagates to -1$\farcs$2, while the red-shifted jet, though initially maintaining the same width, becomes a broad bow-shock type structure at 1$\farcs$2. The bow-shock location appears to coincide with the location of reported H-band emission from 1$\farcs$1 to 1$\farcs$7 to the northeast of DO Tau \citep{Itoh2008}.


Gaussian fitting to intensity profiles allowed extraction of the jet axis position and jet width. In Figure \ref{fig:morphology}, the jet axis position is over-plotted with black dots. There is a clear wiggling of both the blue- and red-shifted jets. The wiggle appears to increase in amplitude with distance from the source. From the wiggle pattern of the blue-shifted jet, we derive an observed (projected) spatial wavelength $\lambda_{\text{obs}}= 90 \pm 5 ~\text{au}$, and semi-opening angle $\theta_{\text{obs}} = 1.3 \pm 0.5 \degr $. While the spatial wavelength is not well constrained on the red-shifted side (with  $\lambda_{\text{obs}} \simeq 100-200~\text{au} $), the wiggles seem to carve out a semi-opening angle of $\theta_{\text{obs}} \simeq  2.6-4.7{\degr}$, which is clearly larger than on the blue-shifted side. The origin of the wiggling will be investigated further in section \ref{sec:discuss_jetwiggle}.

\begin{figure*}
\centering
\includegraphics[width=0.7\textwidth, angle=0]{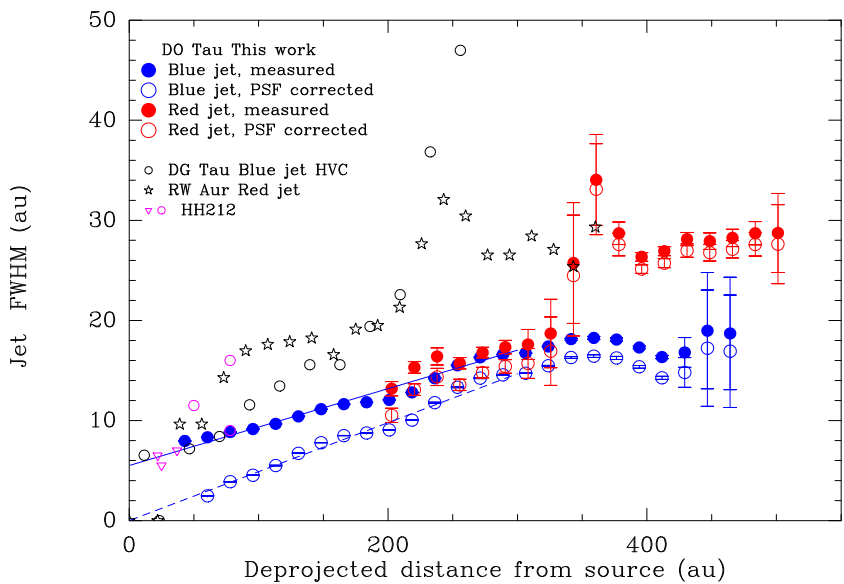}
\caption{Jet FWHM versus deprojected distance from the source. Full circles show the DO Tau FWHM directly measured on the deconvolved images (upper limit to jet diameter) while open circles give the same values corrected for the estimated PSF FWHM (=0$\farcs$06), giving a lower limit to the jet diameter. All measurements from the other jets have been corrected for their respective estimated PSF FWHM. Errors bars for DO Tau are 3$\sigma$.  For DO Tau, we assume an inclination of the jet axis to the line of sight of i=23.5$^{\circ}$. The full and dashed lines are fits of the jet FWHM for z=35-300~au with deprojected semi-opening angles of 1.1$^{\circ}$ and 1.4$^{\circ}$ respectively. We also show FWHM for the DG Tau [Fe {\sc ii}] HVC (blue lobe) from \citet{Agra-Amboage2011},  for the RW Aur jet (red lobe) from  \cite{Woitas2002}, and for the HH~212 jet from \citealp{Lee2017}. Adopted jet inclination to the line of sight are i=37$^{\circ}$ for DG Tau \citep{Bacciotti2018}, i=55.5$^{\circ}$ for RW Aur \citep{Rodriguez2018}.}. 
\label{fig:fwhm} 
\end{figure*}

Figure~\ref{fig:fwhm} shows the jet widths for both jet and counter-jet as a function of deprojected distance from the source. Full symbols show the jet widths directly measured on the deconvolved images, while open symbols show the same values deconvolved from the PSF width, taking the first measure as an estimate of the PSF FWHM. These values give upper and lower limits respectively on the intrinsic jet width. Indeed, since the deconvolution is performed on the continuum subtracted images, no precise knowledge of the PSF in the final deconvolved images is available. An upper limit on the PSF width is obtained by assuming that the jet width is not resolved at distances closest to the source.
Halving the FWHM values, we obtain estimates of the jet radius.
The blue-shifted jet reveals a radius of $r~\le~4~{\rm au}$ at distances from the star of $z=40~{\rm au}$. This increases steadily to $r = 8-9~{\rm au}$ at $z=300~{\rm au}$. Both jets maintain the same jet width in their inner regions, which steadily increases with distance. However, the red-shifted jet reveals an increased width at the location of its apparent bow-shock feature as it propagates from $z=$300~to~400~${\rm au}$. 

Comparing to the literature (figure \ref{fig:fwhm}), it appears that the DO Tau blue-shifted jet is very narrow and strongly collimated with respect to other Class II T Tauri jet targets. Indeed, it is similar to the well-known narrow jet of the very active T Tauri star, RW Aur \citep{Woitas2002}, and to the higher velocity component of the jet from DG Tau \citep{Agra-Amboage2011}.  We derive a semi-opening angle $\simeq$ 1.1-1.4$^{\circ}$ for the blue-shifted jet, for deprojected distances along the jet axis z$_0$ = 40$~-$300~au. No significant change in opening angle is observed over these distances. Thus, our results confirm previous claims that jets already achieve strong collimation on scales of a few tens of au after launching has occurred \citep{Ray2007}. Furthermore, the DO~Tau blue-shifted jet appears very narrow at its base, with an upper limit on its radius of only r $<$ 4 au for a deprojected distance along the jet axis z$_0$=35~au from the star. This upper limit is very similar to the collimation of the base of the HH\,212 Class 0 jet seen with ALMA \citep{Lee2017}. Finally, the derived value of the blue-shifted jet semi-opening angle ($\leq 1.4^{\circ}$) is significantly smaller than the estimated Mach semi-opening angle, $c_s/V_{jet}$ = 4.4$^{\circ}$ assuming $T=8000~K$, indicating the necessity of an additional collimation agent. 
These observations provide direct confirmation that atomic jets originate from the very inner au of the star-disk system, as originally shown by \citet{Hartigan2004}. We also note that the collimation is similar for the inner portions of both lobes of the bipolar jet, in spite of a strong jet velocity asymmetry. 

\subsection{Jet Kinematics} 
\label{sec:results_kinematics}

\subsubsection{Bipolar velocity asymmetry}
\label{sec:results_asymmetry}

\begin{figure}
\centering 
\includegraphics[width=0.4\textwidth]{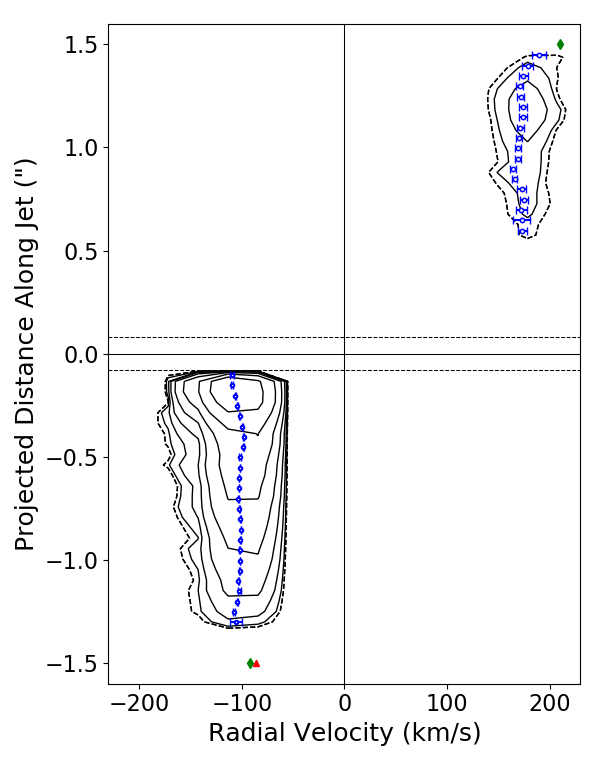}
   \caption{Position-velocity diagram of the DO Tau bipolar jet in [\ion{Fe}{II}] 1.64 $\mu$m~emission along PA=160$^{\circ}$ in a pseudo-slit of width $0\farcs3$. The contour floor is $3 \sigma$ (solid line) and increases in log intervals of 0.3. Note that $2 \sigma$ is also marked (dashed line), and $1 \sigma = 3.5 \times 10^{-16} \text{erg~s}^{-1} \text{cm}^{-2} \text{\AA}^{-1}$. The horizontal solid line marks the star position. The horizontal dashed lines mark the spatial resolution. The blue dashed line shows the variation of the peak velocity centroids, derived from Gaussian fitting to the line profiles. The green diamonds represent the centroid velocities found by \citet{Hirth1997}; red triangle represents the velocity found by \citet{Giannini2019}.}
   \label{pv} 
\end{figure} 

Radial velocities were measured by Gaussian fitting the spectrally binned [\ion{Fe}{II}] 1.64 $\mu$m line at each spatial position. Figure~\ref{pv} shows a position-velocity (PV) diagram of the bright blue-shifted jet and dimmer red-shifted counter-jet. 

Radial velocities are measured to be -103 $\pm$ 1 km~s$^{-1}$ and +173 $\pm$ 4 km~s$^{-1}$ for the jet and counter-jet respectively, when we average over the entire length of the jet. A similar radial velocity asymmetry of -92 km~s$^{-1}$ and +210 km~s$^{-1}$, at 1$\farcs$5 along the jet and counter-jet, was previously reported by \citet{Hirth1997} - see datapoints overplotted in Figure \ref{pv}. More recent results for the blue-shifted jet include \citet{Giannini2019} who report well-resolved low and high velocity components (which remain spectrally unresolved in our PV diagram), with the latter dominating the emission, and having peak velocity $\sim$ -102 and -87 km~s$^{-1}$ for [\ion{O}{I}]$\lambda$6300 and [\ion{Fe}{II}]$\lambda$16440 respectively. \citet{Simon2016} report similar two-component results with centroid velocities of $\sim$ -97 km~s$^{-1}$ for the [\ion{O}{1}]$\lambda$6300 higher velocity component. 
Comparing all these measurements, we see that each species traces a slightly different velocity, but within 10-15~\kms and so no time variability is apparent across epochs. Thus, our measurements confirm previous reports in the literature of the jet velocity asymmetry, and show a limited time variability $\leq 10\%$ in the blue-shifted jet velocity over a 20 year interval, and $\leq 20\%$ in the red-shifted jet velocity.



The striking velocity asymmetry between the jet and counter-jet may be caused by an interaction with the environment as the jet propagates, or it may be intrinsic to the jet launching mechanism (see below). In our data, the velocity asymmetry is observed in all positions where both jets are observed, including as close as $z_{\rm proj}$ =0.5$^{\prime\prime}$ (deprojected distances of 140~au). Furthermore, although the red-shifted jet is not detected closer to the star (likely due to obscuration by the dusty disk), the blue-shifted jet reaches its terminal velocity even closer, at $z_{\rm proj} \le 0.1^{\prime\prime}$, corresponding to deprojected distances of z=28~au. These two results strongly support a velocity asymmetry originating in the launching process itself, rather than resulting from an interaction with the environment. 

\begin{figure}
    \centering
    \includegraphics[width=0.49\textwidth]{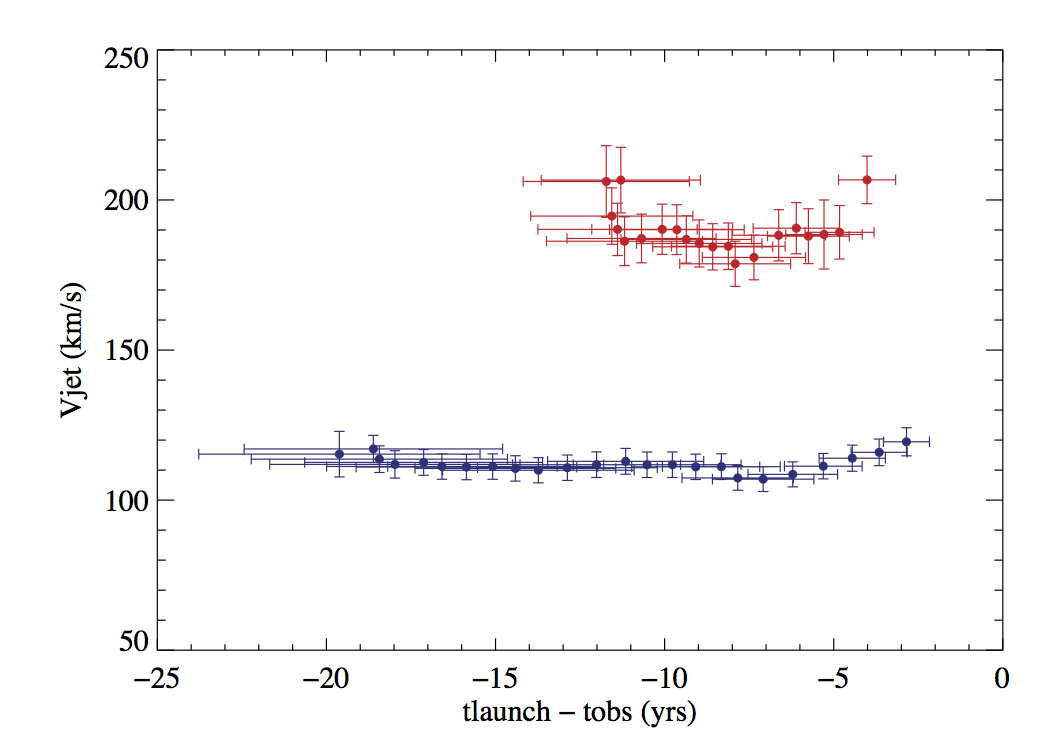}
    \caption{Jet deprojected velocities as a function of their time of launch (measured from the average observation date t$_{\rm obs}$=2009.97) assuming ballistic motion at constant velocity (see text for more details). Error bars include uncertainties on the jet inclination and centroid line of sight velocities.}
    \label{fig:vrad}
\end{figure}

At each projected distance along the jet, $z_{\text{proj}}$, we compute the launch time of the gas parcel of line of sight velocity $V_{\text{rad}} (z_{\text{proj}})$ assuming constant ballistic motion since origin: $t_{\text{launch}} - t_{\text{obs}} = z_{\text{proj}}/(V_{\text{rad}} \times \tan(\text{i}))$, where $t_{\text{obs}}$ is 2009.97 (average of the two observing epochs). Figure~\ref{fig:vrad} shows the deprojected velocities as a function of their time of launch for both jets. This figure represents a reconstructed history of ejection velocities. The plot has a time resolution, $\Delta t = \Delta z /(V_{\text{rad}} \times \tan(\text{i}))$,  on the order of 0.7 years on the blue-shifted side, and 0.4 years on the red-shifted side. The non-detection of the red-shifted jet emission, which would have been launched over the past 4 years, may be due to extinction by the circumstellar disk, as per section \ref{sec:results_morph}. The blue-shifted jet shows an average deprojected velocity of 112 $\pm$ 3 \kms over the past 20~years, while the red-shifted jet shows an average deprojected velocity of 190 $\pm$ 8 \kms over 8~years. Launching of both jets co-existed for at least 8~years, and the velocity asymmetry (with an almost constant factor of $\simeq$ 1.7) is sustained over at least that period of time. Since we concluded that the velocity asymmetry is likely to originate in the launching process, the duration of the sustained asymmetry provides a useful constraint on jet launch models (see section \ref{sec:discuss_asymmetry}). 

\subsubsection{Jet shocks}
\label{sec:results_shocks}

While we do not identify any great velocity variations over time (thus lending a stability to the velocity asymmetry of the bipolar jet), the fact that our data is spatially resolved allows a snapshot of velocity variations within the jet at a fixed point in time. Considering velocity variations {\em along} the jet (for this epoch), Figure \ref{fig:velocity_map_jet} shows a radial velocity map of the bipolar jet, with an associated error map. We can see that the blue-shifted jet has a radial velocity of -110 km~s$^{-1}$ at 0$\farcs$1 from the star which decreases to -90 km~s$^{-1}$ at 0$\farcs$4, before increasing again at 0$\farcs$5 to -100 km~s$^{-1}$ and more or less maintaining this velocity further downstream. Such changes in velocity are expected at shock fronts, where the gas is collisionally excited. However, we do not clearly identify emission knots in the blue-shifted jet, and these 20 km~s$^{-1}$ variations in velocity do not correspond to any measurable change in intensity or morphological feature at this distance, recall Figure \ref{fig:morphology}. Intriguingly, the initial decrease in velocity occurs over similar distances as a jump in electron number density (see section \ref{sec:results_n_e}). However, the gradient in velocity is significantly smoother. This behaviour is not compatible with a shock as the velocity would be expected to drop sharply at the location of the shock front, which is not observed here. 

\subsubsection{Jet rotation}
\label{sec:results_rotation}

We consider velocity variations {\em across} the jet, which may indicate a rotation of the flow or alternatively may be caused by a wiggling of the jet. We first note that the borders of the jet travel at a lower velocity to the on-axis component, dropping by $\sim$ 5 km~s$^{-1}$. This reflects the onion-like layered velocity structure observed in other high resolution case studies (e.g. \cite{Bacciotti2000}). At a distance larger than 0$\farcs$6 along the blue-shifted jet, the SE border of the jet appears to be more red-shifted. These distances coincide with the location of a wiggle in the jet axis. Meanwhile, at the same distances along the red-shifted jet, the SE border is more blue-shifted.

\begin{figure*}
\centering
\includegraphics[height=8cm]{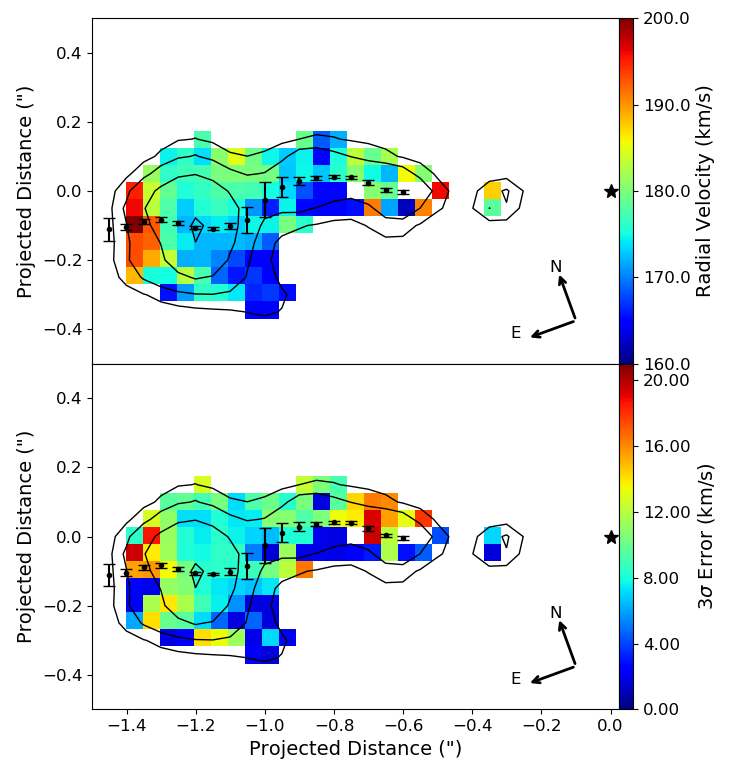}
\includegraphics[height=8cm]{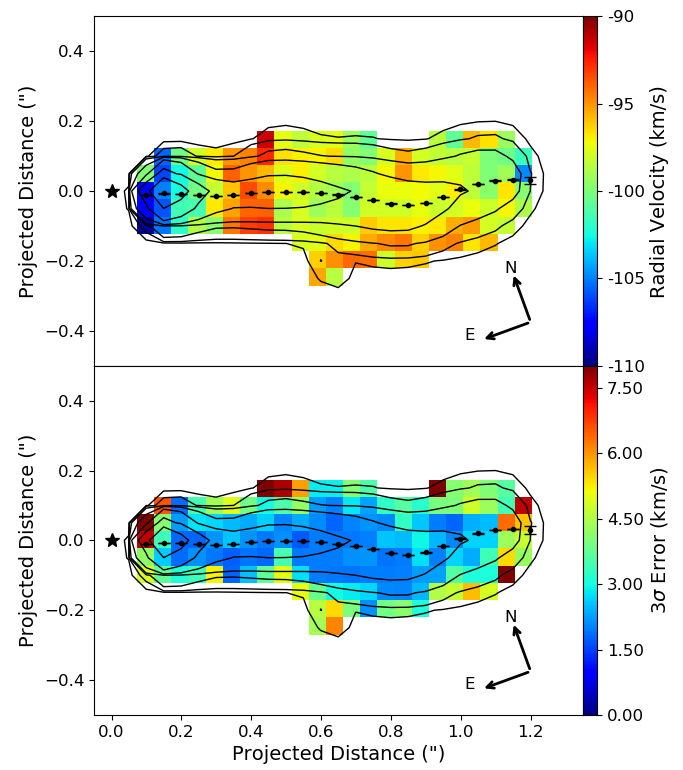}
\caption{Centroid velocity map of the DO Tau bipolar jet, derived from Gaussian fitting the [\ion{Fe}{II}] 1.64 $\mu$m line profile at each spaxel (see text for more details). Brightness contours from figure \ref{fig:morphology} are overlaid, along with the jet axis centroids. 3$\sigma$ errors are given in the bottom panels. }
   \label{fig:velocity_map_jet}
\end{figure*}




A closer examination of a possible asymmetry in radial velocity across the jet first requires identification of the jet axis position. At a given distance from the star, the jet axis position was identified by spectrally binning the [\ion{Fe}{II}] 1.64 emission line and fitting a Gaussian profile to determine the peak centroid. The jet radial velocity was measured at each transverse position from the jet axis, via Gaussian profile fitting.
Measurements equidistant from the axis were compared to identify differences in radial velocity. Systematic asymmetries may be interpreted as a signature of a rotation of the flow. A comparison with the disk rotation sense, if possible, is a fundamental consistency check on the interpretation, i.e. the disk and both lobes of the bipolar jet should all be seen to rotate in the same direction. 

\begin{figure*}
\centering
    \includegraphics[width=0.33\textwidth]{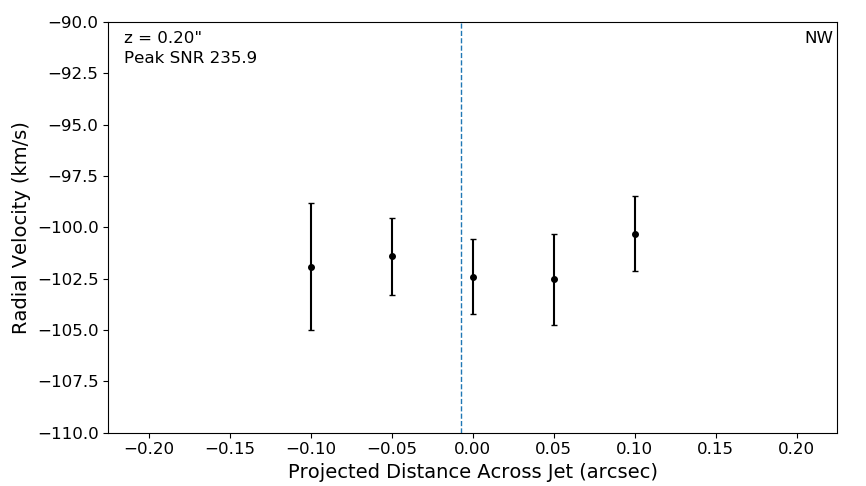}
    \includegraphics[width=0.33\textwidth]{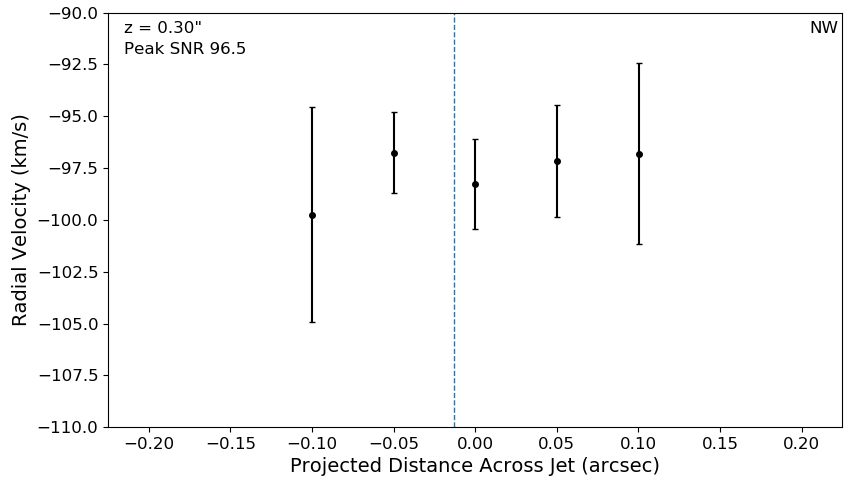}
    \includegraphics[width=0.33\textwidth]{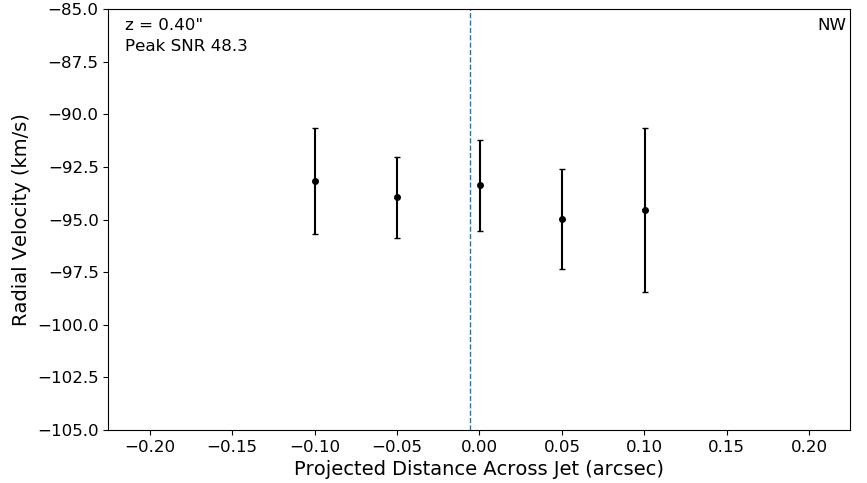}
    \includegraphics[width=0.33\textwidth]{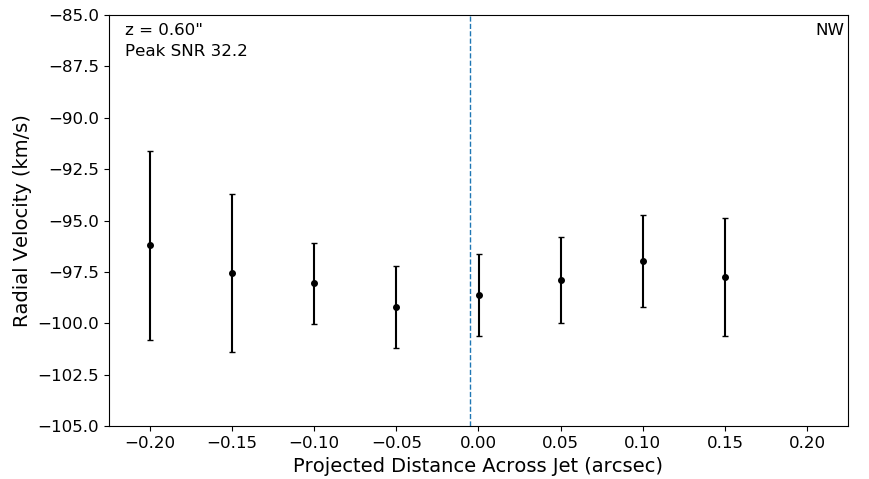}
    \includegraphics[width=0.33\textwidth]{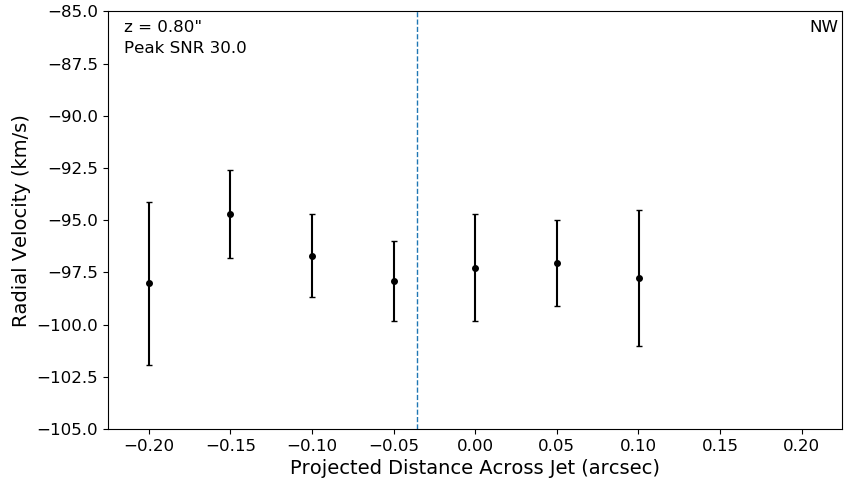}
    \includegraphics[width=0.33\textwidth]{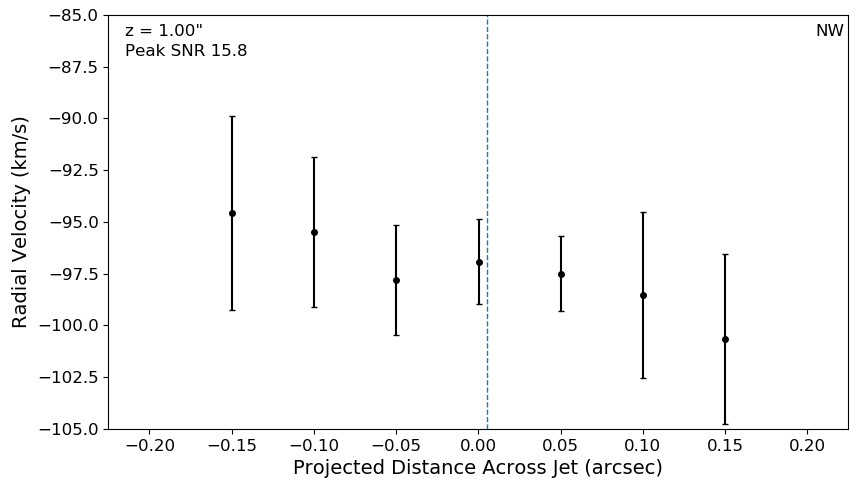}
    \caption{Radial velocity profiles across the jet at various distances, z, from the star. The vertical line marks the measured jet axis position at this distance z, whereas the zero position on the x-axis marks the nominal jet axis position according to the jet PA. The compass orientation of the x-axis is from SE to NW. A jet rotation signature should reveal itself as a systematic gradient in radial velocities across the jet in a sense that matches the disk rotation sense. (Recall the kinematics were measured on data before PSF deconvolution i.e. PSF FWHM = 0$"$15.) No systematic gradient is observed. Errors are 3$\sigma$.} 
    \label{fig:velocity_trans}
\end{figure*}

Figure \ref{fig:velocity_trans} shows the transverse radial velocity profile at a selection of projected distances, z, from the star. In the inner jet region ($< 0\farcs5$), we would expect that the NW side of the jet is less blue-shifted (redder), as this would match the sense of disk rotation, Figure \ref{fig:morphology}. However, it is clear that we cannot identify a gradient in the Doppler profile, between positions symmetric about the jet axis, as would be expected if due to rotation, due to the large error bars. Since errors on centroid velocity fits are signal-to-noise dependant, measurements in the borders of the jet suffer. This hampers our ability to detect small differences in radial velocity between the jet borders. 

Meanwhile, recall in figure \ref{fig:velocity_map_jet} that from 0$\farcs$6 along the jet, the SE border of the approaching jet seems more red-shifted. While this does not agree with the disk rotation sense, the redder velocities coincide with the location of a wiggle in the jet axis. By contrast, the receding jet shows bluer velocities in the SE border, in agreement with the disk rotation sense. Unfortunately, error bars on jet velocities, as well as a wiggle of the jet axis, forbid certainty in any claim to observe jet rotation. We can, however, provide an upper limit on a possible rotation signature based on our calculated error bars (see section \ref{sec:discuss_footpoint}).

\subsection{Electron Number Density of the Jet} 
\label{sec:results_n_e}

\begin{figure}
    \centering
    \includegraphics[width=0.48\textwidth]{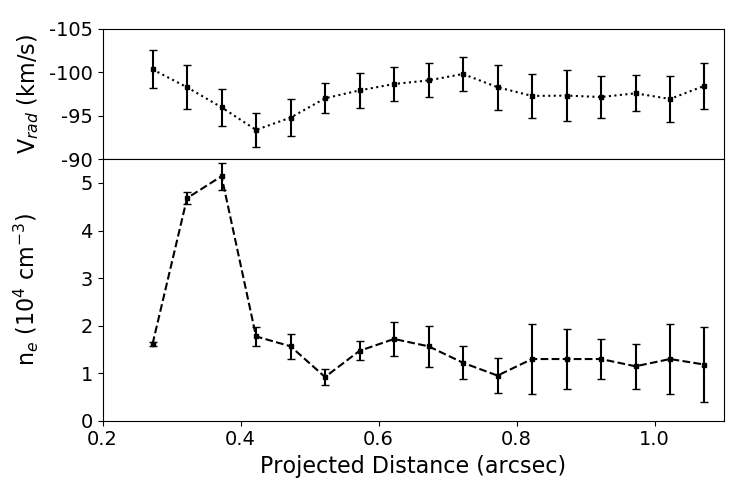}
    \caption{Top: Blue-shifted jet centroid velocities with 3$\sigma$ errors. Bottom: Electron number density, $n_e$, derived from the line ratio [\ion{Fe}{II}] 1.53$\mu$m/1.64$\mu$m at various distances along the blue-shifted jet. [\ion{Fe}{II}] 1.53$\mu$m emission close to the source was contaminated by residuals from continuum subtraction.} 
    \label{fig:electron_density} 
\end{figure}

A direct measure of the electron number density, $n_{e}$, can be obtained from the emission line ratio [\ion{Fe}{II}] 1.53$\mu$m/1.64$\mu$m \citep{Pradhan1993}. 
To increase signal-to-noise, emission line fluxes were binned spatially across the jet. This allowed a 1D plot of flux ratio along the jet axis from which the electron number density was derived. Unfortunately, the [\ion{Fe}{II}]~1.53$\mu$m emission for the red-shifted jet was too faint, and so the electron number density could not be estimated here. 
Figure \ref{fig:electron_density} shows estimates for the blue-shifted jet. We caution that our $n_e$ measurement closest to the star may still be contaminated by photospheric subtraction residuals. We see that derived $n_{e}$ values range from $1~\text{to}~5 \times 10^4$~cm$^{-3}$ close to the star, and drop to a plateau at $\simeq 10^4$~cm$^{-3}$ beyond projected distances along the jet of z=0.4-1$^{\prime\prime}$. These values are lower than the $n_{e}$ values of 10$^{4.5}$ to  10$^{5}$~cm$^{-3}$, derived by \citet{Giannini2019}, from a spectrum on source with seeing of 0$\farcs$8 (i.e. not spatially resolved along the jet). \citet{Giannini2019} values are likely strongly dominated by the very inner denser regions of the jets, which we do not probe here due to strong continuum residuals. Hence, it is likely that $n_e$ remains high closer to the star, i.e. our first datapoint may be discounted. Such high values are also typically found at the base of other T Tauri jets (see e.g. \citealp{Coffey2008}). 





\subsection{Bipolar jet mass flux} 
\label{sec:results_m_dot}

A critical parameter in modelling jet physics is the mass flux of the jet. Using our measurements, the jet mass flux was estimated with two methods, described in detail in \citet{Nisini2005} and  \citet{Agra-Amboage2011}. 

Method 1: For a jet uniformly filled with a number density of hydrogen nuclei, n$_H$, the mass flux is given by \citep[see Eq. 2 in][]{Agra-Amboage2011}: 
\begin{equation}
\begin{split}
    \dot{\text{M}}_{J} = 1.3 \times 10^{-9} \text{M}_{\odot}~\text{yr}^{-1} \left( \frac{\text{n}_H}{10^5~\text{cm}^{-3}} \right) \\ 
    \times \left( \frac{\text{D}_J}{14~\text{au}} \right)^2 \left( \frac{\text{V}_J}{100~\text{km} \text{s}^{-1}} \right)
\end{split}
\end{equation}
where 
D$_{J}$ is the jet diameter and V$_J$ the jet velocity. The electron number density $n_e$ and ionisation fraction $x_e$ together allow an estimate of the hydrogen number density, $\text{n}_H=\text{n}_e/\text{x}_e$. For DO Tau blue-shifted jet, the ionisation fraction, $\text{x}_e$, was reported as 0.01 at velocities of -100 km~s$^{-1}$ from high spectral resolution data \citep[see Fig. 10][]{Giannini2019}. Since there is no spatial information (i.e. how the ionisation fraction changes with distance along the jet), we adopt this value for all positions along the blue-shifted jet. The jet diameter is taken as the jet FWHM directly measured in the deconvolved maps. This gives an upper limit on the jet diameter (in the sense that we we do not correct for the PSF FWHM). 
The deprojected jet velocity is given by $\text{V}_{J} = \text{V}_r/\text{cos}(i_{jet})$, where $\text{i}_{jet}$ is the jet inclination angle to the line of sight, and $\text{V}_r$ is the radial velocity of the jet. The deprojected blue-shifted jet velocity is thus $\text{V}_{J} = -109$~km~s$^{-1}$. 
Method 1 gives an upper limit on the mass flux. Indeed, in addition to the upper limit on the jet radius, we also likely over-estimate the hydrogen number density, given that we assume a constant ionisation along the jet whereas the ionisation may vary (e.g. \citealp{Bacciotti1999}). Indeed, the value taken from \citet{Giannini2019} is probably dominated by the inner brighter regions close to the star, and so better represents the base of the jet. 

Method 2 allows us to derive an estimate of the jet mass flux from the [\ion{Fe}{II}] line luminosity and electronic density estimates. This method assumes optically thin emission, uniform excitation conditions within the emitting volume and iron fully singly ionized (Fe+/Fe=1). The [\ion{Fe}{II}] line luminosity per unit length can then be directly related to the mass flux with the following equation \citep[eq. 4 in ][]{Agra-Amboage2011}:
\begin{equation}
\begin{split}
\dot{\text{M}}_{J} =  1.45 \times 10^{-8}~ \text{M}_{\odot}~\text{yr}^{-1} \left(1 + \frac{3.5 \times 10^4}{\text{n}_e (\text{cm}^{-3})} \right) \left( \frac{\text{L}_{[\text{Fe}{\sc ii}]}}{10^{-4} ~\text{L}_{\odot}} \right) \\ \times \left( \frac{\text{V}_t}{150~\text{km} \text{s}^{-1}} \right) \left( \frac{\text{l}_t}{2  10^{15} \text{cm}} \right)^{-1}~\left( \frac{[\text{Fe}/\text{H}]}{[\text{Fe}/\text{H}]_{\odot}} \right)^{-1} 
\end{split}
\end{equation}
where V$_t$ is the jet tangential velocity projected on the plane of the sky, and l$_t$  the length of the aperture both projected onto the plane of the sky. We take v$_t$=43.5 \kms (calculated from $i=$23.5$^{\circ}$ and $v_r$=-103 \kms) and l$_t$ = 0$\farcs$05 = 10$^{14}$ cm (at 140 pc). This method also depends on the Fe gas phase abundance. \citet{Giannini2019} has shown that the Fe gas phase abundance is close to solar in the DO Tau blue-shifted jet at higher velocities, with [Fe/H] $\simeq$ 70\% [Fe/H]$_{\sun}$. No spatial information is available for the evolution of this ratio along the jet so we take here a ratio value of 1 at all positions along the jet. No correction for extinction was included either, since the already low visual extinction towards DO Tau, A$_v$= 0.75, has even less impact in the near-IR. 
Method 2 gives a lower limit on the mass flux. Here, a lack of correction for extinction leads to an under-estimated luminosity. Also, we do not correct for iron depletion, which again yields a lower limit on the mass flux. 

\begin{figure}
    \centering
    \includegraphics[width=0.49\textwidth]{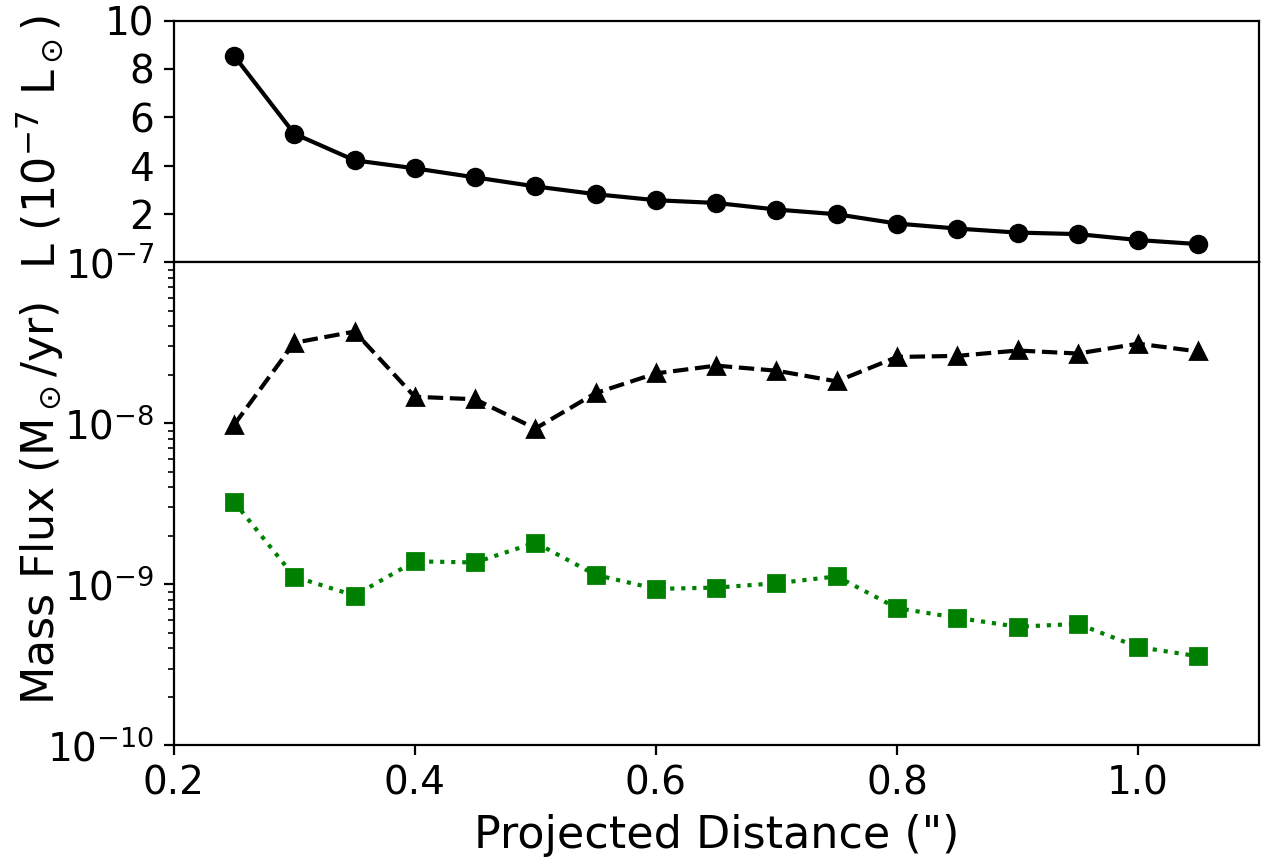}
    \caption{Top panel shows the [\ion{Fe}{II}] 1.64$\mu$m line luminosity per unit length for the blue-shifted jet. (The red-shifted jet has a typical value of $\sim$ 3 $\times$ 10$^{-8}$ L${\sun}$ per unit length) Bottom panel shows the jet mass flux calculated using two methods: method 1 (black) based on n$_H$ and jet radius estimate, method 2 (green) based on the [\ion{Fe}{II}] 1.64 $\mu$m line luminosity per unit length and the electronic density n$_e$ to find the emitting mass. Both methods are described in more details in the text. Method 1 gives an upper limit, while method 2 is a lower limit.} 
    \label{fig:mass_loss}
\end{figure}

Figure \ref{fig:mass_loss} shows our results for the derived mass flux along the blue-shifted jet. Errors are on the order of 30\%, increasing to $>$ 50\% beyond 0$\farcs$9 from the source. 
Since Method 1 provides an upper limit and Method 2 a lower limit to the mass flux, we take an average of the two methods.  
Note that the mass flux values presented in Figure \ref{fig:mass_loss} rely on a one-dimensional plot of the electron number density along the jet (see Figure \ref{fig:electron_density}). Given the faintness of the 1.53~$\mu$m line, it was not possible to produce a 2D map of the density. Additionally, the total density is not accessible without a measure of the ionisation fraction for which we have used an estimate measured by \citet{Giannini2019} near the star (since they do not spatially resolve the jet). Note that, with the assumption of a constant ionisation fraction, the variation in mass flux with distance from the star reflects the variation in electron number density. However, we caution that spatially resolved measurements of ionisation fraction in other Class~II jets show that this quantity can fluctuate spatially by a factor of ten within a few 100 au of the source  \citep[e.g.][]{LavalleyFouquet2000, Bacciotti2002, Hartigan2007, Maurri2014}. Therefore, we cannot exclude that the fluctuations in Figure \ref{fig:mass_loss} could be caused by underlying spatial variations in ionisation fraction, rather than a true change in mass flux. So only the average value of mass flux will be considered hereafter. Thus, we estimate 5 $\times$ 10$^{-9}$ M$_{\odot}$ yr$^{-1}$ for the blue-shifted jet.


When considering the red-shifted jet, we note that the [Fe {\sc ii}] 1.64~$\mu$m line is about ten times fainter than in the blue-shifted jet, and the 1.53~$\mu$m line is not detected in the red-shifted jet in our data. Instead, to estimate the mass flux in the red-shifted jet, we use the value for electron density found by \citet{Hirth1997}. In the red-shifted jet, an average electron density of 5000~cm$^{-3}$ is found between 2-4$\arcsec$ from the star. Assuming the ionisation fraction is similar, this suggests a possible mass flux on the order of 10$^{-8}$ M$_{\odot}$ yr$^{-1}$ in the red-shifted jet. The mass loss rates compare well with typical values derived for similar high accretion targets. In particular, RW Aur reveals 2.6 and 2 $\times$ 10$^{-9}$ M$_{\odot}$ yr$^{-1}$ for the red- and blue-shifted jet, respectively \citep{Melnikov2009}.

We calculate a mass accretion to ejection ratio by using the mass accretion rate, $\dot{M}_{acc}$ $\ge$ 1.44 $\times$ 10$^{-7}$ $M_{\sun} yr^{-1}$ \citep{Gullbring1998}. Summing our mass flux estimate for the red- and blue- shifted jets, and comparing this to the mass accretion rate, we obtain a ratio of 0.10. Alternatively, since our red-shifted jet mass flux estimate is less accurate, we assume that the red-shifted jet has the same mass flux as the blue-shifted jet (an assumption for asymmetric jets supported by \cite{Melnikov2009}, we double our one-sided value of $\dot{M}_{jet}$ $\sim$ 5$\times$ 10$^{-9}$ M$_{\odot}$ yr$^{-1}$ to achieve a ratio for the bipolar jet of 2$\dot{M}_{jet}$/$\dot{M}_{acc}$ $\sim$ 0.07. 
These ratios are in line with literature values for the HVC component in similar targets, such as e.g. DG Tau \citep{Maurri2014}, RY~Tau \citep{Agra-Amboage2011}, RW Aur \citep{Melnikov2009} and compatible with predictions from MHD launching models \citep{Ferreira2006}. 

\section{Discussion} 
\label{sec:discussion}

\subsection{The Origin of the Bipolar Jet}
\label{sec:discuss_footpoint} 

Assuming the magneto-centrifugal mechanism of jet launching, our observations can potentially provide a way of differentiating between competing models. The main long-standing contenders in this class of model are: accretion powered stellar winds launched from open magnetic field lines anchored at the stellar surface \citep{Matt2005}; winds launched from the interaction of the stellar magnetosphere and inner disk from r $\sim$ 0.05 au \citep{Shu2000,Romanova2009,Zanni2013}; disk-winds launched from a range of radii in the disk \citep{Pudritz2007, Ferreira1997}. 

Our high resolution observations of the jet base allowed us to probe a possible jet rotation signature, which can be used to indicate the launch-point of the jet. Unfortunately, the size of the error bars on radial velocities measured across the jet prevent any claim of a jet rotation detection. We can nevertheless impose an upper limit on any rotation present. The errors in our measurements allow us to provide an upper limit on the jet toroidal velocity, and hence provide constraints on jet launching models. 

However, these models critically rely on an assumption of a steady, axisymmetric flow. Clearly, evidence of jet wiggling may call into question this assumption. Indeed, \citet{Cerqueira2006} showed that a precessing and variable velocity jet can present transverse asymmetries in velocities similar to rotation signatures. However, these simulations include strong velocity variability at the source (with $\Delta V = \pm 100$ \kms). The observed line of sight velocity variations along the DO Tau jets do not exceed $\pm 5 $\kms in the blue, $\pm 13$\kms in the red, in the central 150 au. In addition, our observations do not show any clear evidence for shock formation (no emission knot detected) or brightness asymmetries across the blue-shifted jet. Therefore, we assume that despite the small wiggling detected (with a deprojected semi-opening angle of 0.5$^{\circ}$ more than twice smaller than the jet beam semi-opening angle of 1.1-1.4$^{\circ}$, see Section \ref{sec:results_morph}), the low level of velocity variability does not produce shocks strong enough to introduce significant departure from axisymmetry for the global flow. No simulations of jet propagation with such conditions exist to our knowledge, and would be required to fully test this hypothesis. For DO Tau, the wiggling introduces possible changes in the inclination angle of the flow axis to the line of sight of $< 0.5^{\circ}$ for the blue-shifted jet, and $< 2.5^{\circ}$ for the red-shifted jet. These variations are smaller than the assumed global uncertainty of $5^{\circ}$ taken for the jet inclination to the line of sight. We therefore apply the method outlined below to derive an upper limit on the launching radius.

Assuming a steady, axisymmetric and cold MHD ejection process, the radius in the disk-plane where the jet originates can be calculated using the method of \citet{Anderson2003}. This method relies on the general conservation principles of total specific energy and momentum along each magnetic surface, thus removing the magnetic term and so allowing a calculation of the launch radius from observed velocities alone. The approximation below is valid when the observed jet poloidal velocity is much greater than the Keplerian velocity at the launch point, which is verified a-posteriori, see below. 
Hence, the jet launch radius in the disk plane, $\bar{r}_{0}$, is given by:

\begin{equation}
\begin{split}
\bar{r}_{0} \simeq 0.7~\text{au} \left(\frac{\,\bar{r}_{\infty}}{10~\text{au}}\right)^{2/3} \left ( \frac{\text{v}_{\phi ,\infty}}{10~\text{km~s$^{-1}$}} \right )^{2/3} \\ \times \left ( \frac{\text{v}_{p ,\infty}}{100 ~\text{km~s$^{-1}$}} \right )^{-4/3} \left ( \frac{\text{M}_{*}}{1~\text{M}_{\odot} } \right )^{1/3}
\end{split}
\end{equation}

\noindent 
where $\bar{r}_{\infty}$ is the radius from the jet axis of the observation (at a distance from the disk plane of effectively infinity, z=$\infty$), v$_{{\phi}{,\infty}}$ is the toroidal velocity observed at z=$\infty$, v$_{{p}{,\infty}}$ is the poloidal velocity observed at z=$\infty$, and M$_{*}$ is the mass of the star. 

The poloidal velocity can be derived from the radial velocity, $\bar{{\text v}}_{\text{r}}$ = $\text{v}_p$ cos($i_{\rm jet}$) while the toroidal velocity can be derived from the difference in radial velocities either side of the jet, $\Delta \text{V}_{\text{r}}$ =  2 v$_{\phi}$ sin($i_{\rm jet}$), where the jet inclination angle is with respect to the line of sight. We make our calculations based on velocity measurements at a distance {\em along} the jet of z$\sim$0$\farcs$4 (corresponding to a deprojected distance z$_{0}$=112~au). We want to observe as close to the jet base as possible, to maximize signal-to-noise and minimise effects of jet wiggling further downstream, but far enough from the base to resolve the jet in the transverse direction (without deconvolution). We take measurements at a distance {\em across} the jet from the jet axis of r = 0$\farcs$1. We avoid using measurements very close to the jet axis because it is expected that, as we draw closer to the jet axis, any radial velocity gradients across the jet will be washed-out by beam convolution effects (\citealp{Pesenti2004};  \citealp{Tabone2020}). 



Referring to Figure \ref{fig:velocity_trans}, we see that for z$\sim$0$\farcs$4 (top right panel) we measure an average radial velocity of -95 km~s$^{-1}$ which gives a jet poloidal velocity, $\text{v}_{p}$ = -103.6~km~s$^{-1}$ $\pm$ 7.5~km~s$^{-1}$. For r = 0$\farcs$1 (outer datapoints), the error bars give us an upper limit on the difference in radial velocity, $\Delta \text{v}_r$ $<$ 5 km~s$^{-1}$. This gives us a 3 $\sigma$ upper limit on toroidal velocity of $\text{v}_{\phi}$ $<$ 6.3 km~s$^{-1}$  $\pm$ 0.7~km~s$^{-1}$. We thus achieve an upper limit on the specific angular momentum, r $\times$ v$_{\phi}$, carried by the jet of $<$ 87~au~km~s$^{-1}$ $\pm$ 19 au~km~s$^{-1}$. For a stellar mass $M_{\star} \sim$ 0.59~$M_{\sun}$ (and assuming an uncertainty of 30\% on the stellar mass), we calculate an upper limit on the launch radius in the disk-plane of $\bar{r}_{0}$ $<$ 0.5~$\pm$ 0.1~au. We perform the same calculation for the red-shifted jet at a position of z$\sim$0$\farcs$8, i.e. before variations in the jet axis position become too large. 
At this position, we find an average radial velocity of 170 km~s$^{-1}$ giving a poloidal velocity of 185~km~s$^{-1}$ $\pm$ 12~km~s$^{-1}$. 
From the upper limit on the difference in radial velocity, $\Delta \text{v}_r$ $<$ 7 km~s$^{-1}$ at a jet radius of r = 0.1$\farcs$, we derive an upper limit on the toroidal velocity $\text{v}_\phi$ $<$ 8.7 km~s$^{-1}$ $\pm$ 0.9~km~s$^{-1}$ and specific angular momentum $r {\rm v}_{\phi}$ $<$ 122~au~km~s$^{-1}$ $\pm$ 27.1~au~km~s$^{-1}$. From these values for the poloidal velocity and specific angular momentum, we infer an upper limit on the launching radius of the red-shifted jet to $\bar{r}_{0}$ $<$ 0.3~au $\pm$ 0.06~au. \textbf{Our derived upper limits on the launching radii differ between the two lobes (r$_{0}$ $<$ 0.5 $\pm$ 0.1~au for the blue lobe, r$_{0}$ $<$ 0.3 $\pm$ 0.06~au for the red lobe). However, since these values are \emph{upper limits}, they do not exclude the same launching radius for both sides of the jet. We note, however, that in each lobe we measure different terminal velocities very close to the source. This points to a possible 
asymmetric launching process which could imply different launch radii for the two jet lobes. Possible origins and effects of an asymmetric jet launch are discussed in Section \ref{sec:discuss_asymmetry}.
}

\begin{figure}
    \centering
    \includegraphics[width=0.45\textwidth]{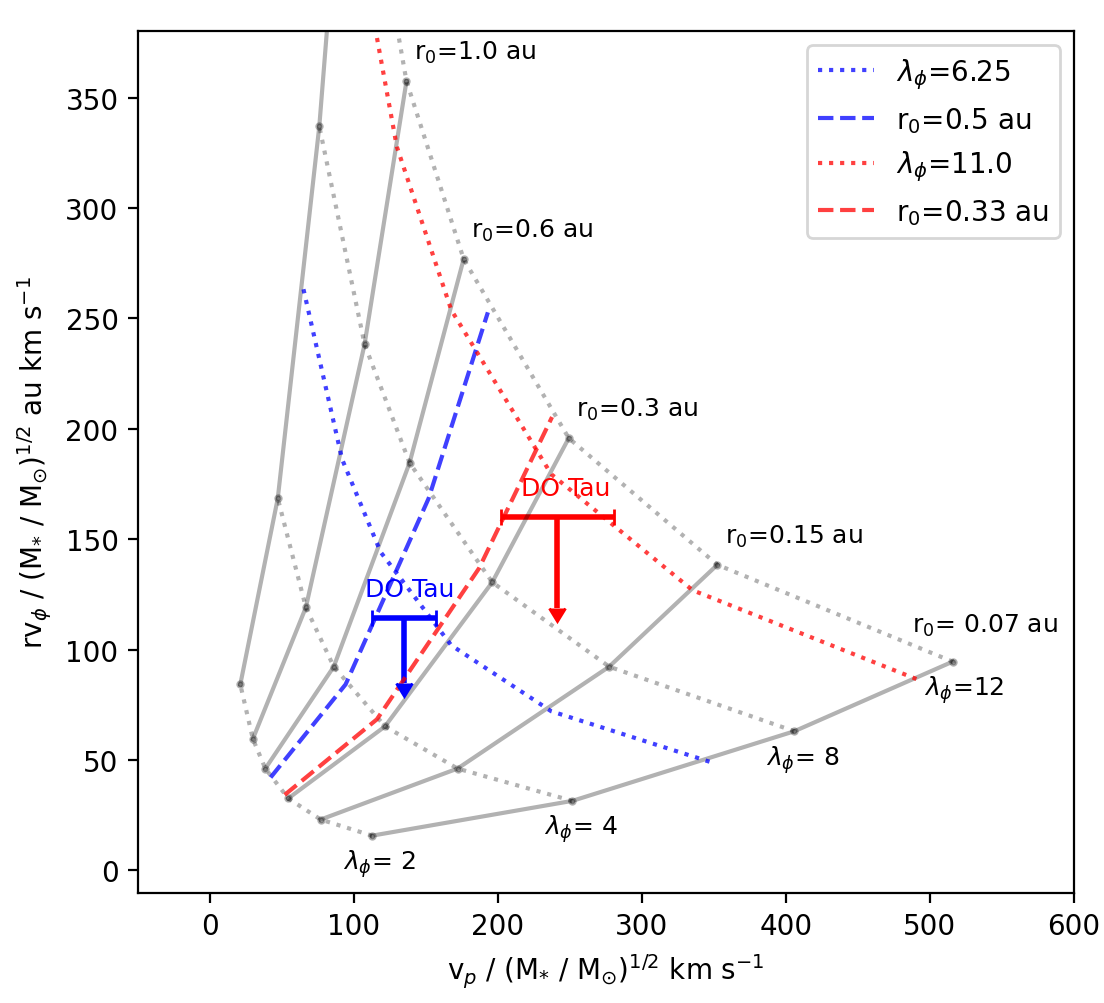}
    \caption{Parameter space of the magnetic lever arm parameter, $\lambda_{\phi}$, and jet launching radius, $r_0$, predicted for steady magneto-centrifugal disk winds, based on the relationship between jet kinematics, $rv_\phi$ and $v_p$ (see \citealp{Ferreira2006}). DO Tau blue- and red-shifted jet kinematics are plotted, with 1$\sigma$ errors, revealing theoretical $\lambda_{\phi}$ and $r_0$ upper limits.} 
    \label{fig:launching} 
\end{figure}

Besides the launch point, $r_0$, another important parameter that can be extracted from observed kinematics to constrain the MHD launching mechanism is the "magnetic lever arm parameter" $\lambda$ (\citealp{Anderson2003} and \citealp{Ferreira2006}). This parameter measures the total specific angular momentum extracted by the wind in units of the Keplerian value at its footpoint in the disk. It may be shown from steady MHD theory that $\lambda$ $\simeq$ ($r_A$/$r_0$)$^2$ where $r_A$ is the Alfv\'{e}n radius\footnote{The Alfv\'{e}n radius of the magnetic streamline is the distance along the outflow at which the accelerating plasma reaches the velocity of the Alfv\'{e}n wave propagating along the flow, $v_A$=(B$_p$/(4$\pi \rho$))$^{1/2}$.}  \citep{BlandfordPayne1982}. Therefore, the asymptotic level of wind rotation is the same as if matter was forced to co-rotate rigidly with the field up to the Alfv\'{e}n radius $r_A$, and conserve its angular momentum beyond this point. The lever arm is, therefore, a measure of the efficiency of the mechanism for extracting angular momentum from the disk. (In reality, however, the conversion of field torsion into outflow rotation takes place over scales $>$ $r_A$ so that the observed $\lambda_{\phi}$ estimated from the specific angular momentum of the matter only gives a lower limit to the true $\lambda$ (\citealp{Pesenti2004}, \citealp{Ferreira2006})). 
Note that $\lambda_{\phi}$ reaches $\lambda$ when {\it all} the magnetic energy is converted to kinetic energy, i.e. when all angular momentum has been transferred to the matter. This typically occurs when $z/r_0 >> 100$ \citep{Tabone2020}. Given the distance at which we make our measurements, z = 0$\farcs$4-0$\farcs$8 ($\sim$ 60-120 au), and the small inferred values of $r_0 <$ 0.3-0.5~au, we would thus expect $\lambda_\phi \sim \lambda$. Using Equation~10 from \cite{Ferreira2006}, we retrieve an upper limit on $\lambda_{\phi}$ independently of $r_0$ from observed values of $rv_\phi$ and $v_p$. (For graphical derivation, see Figure~\ref{fig:launching}.) We find $\lambda_{\phi} \sim \lambda \leq$ 6.25 and 11, for the blue- and red-shifted jets respectively. 



Our derived upper limit for the DO Tau blue-shifted launching radius is consistent with the low Fe gas phase depletion observed in this jet \citep{Giannini2019}, suggesting that the Fe jet originates from close to the dust sublimation radius in the disk, estimated to be $\sim$ 0.13~au by \citet{Eisner2014}. Our derived upper limit on $\bar{r}_{0}$ also matches the computed upper limits of $r_0$ = 0.44 and 1.6~au for the RW~Aur blue- and red-shifted jet respectively \citep{Woitas2005}, and of $r_0 < 0.45$~au for the RY~Tau blue-shifted jet \citep{Coffey2015}. Moderate $\lambda$ values such as the one derived here have been also favoured for the DG~Tau jet \citep{Pesenti2004}. They correspond to warm MHD disk-wind solutions such as the ones computed by \citep{Casse2000}.

Our results do not allow discrimination between the competing launching mechanisms. However, they confirm that the high-velocity jets originate from, at most, 0.5~au $\pm$ 0.1~au from the star in the disk plane. 
We caution however that these results only relate to the very narrow jet component which is traced by the high velocity [\ion{Fe}{II}] line. It is possible that there co-exists a lower velocity, broader outflow component which is launched from larger radii in the disk-plane.  Indeed, a low-velocity component is also detected in DO~Tau in some typical jet emission lines \citep{Giannini2019}. These authors have shown that physical properties (excitation temperature, densities, ionisation fractions) vary continuously with velocities across the line profiles suggesting a common physical origin for both the lower and higher velocity components. Furthermore, \citet{FernandezLopez2020} estimate a larger upper limit of $\le$~15 au on the launching radius for the wind component driving the ringed CO outflow which they recently identified. 

\subsection{The origin of the velocity asymmetry}
\label{sec:discuss_asymmetry} 

In section \ref{sec:results_asymmetry}, we report that the velocity asymmetry between the DO~Tau blue- and red-shifted jets is traced close to the source, indicating that it is likely to be caused by the jet launch mechanism. 
We also report that the asymmetry is constant at a factor 1.7 and sustained for around 7~years. For a typical inner disk truncation radius of 0.05~au, the Keplerian orbital timescale at the inner disk around a 0.59~M$_{\sun}$ star is 0.014~yrs ($\simeq$ 5~days). Therefore, the velocity asymmetry in the DO Tau bipolar jet is sustained over a duration of at least 500 times the orbital period of the inner disk. We investigate below the constraints brought by these results on asymmetric jet launching models. 



\citet{Lovelace2010}, \citet{Dyda2015} and \citet{Lii2014} have all studied the launching of asymmetric outflows from the magnetospheres of rapidly rotating stars in the propeller regime. They find that the presence of a small dipole component leads to the formation of one-sided accretion funnel flows accompanied by episodic outflows in the opposite hemisphere. 
The dominant direction of ejection changes stochastically on timescales of $\simeq$ 30-50~inner disk orbital periods.
The inclusion of a disk magnetic field does not significantly affect these results \citep{Dyda2015,Lii2014}. On longer timescales, the jets will therefore look essentially symmetric.
Such models therefore cannot account for the velocity asymmetry sustained over typically an order of magnitude longer timescales observed in the DO Tau bipolar jet. 

 

A more promising scenario is provided by recent numerical works studying the launching of outflows from magnetised disks. If originating from an MHD disk-wind, a difference of a factor 4 in launching radius would account for a difference of a factor 2 in terminal velocities. Alternatively, different lambda values could be achieved between the two surfaces of the disk. The pioneering studies of \citet{Fendt2013} and \citet{Dyda2015}, using a simplified $\alpha$ prescription for the disk viscosity, have shown that indeed asymmetric high velocity jets can develop and endure over long lasting timescales, $\simeq$ a few hundred orbital timescales at the inner disk radius, depending on the disk magnetic properties. Bipolar jet velocity asymmetries of up to a factor 2-2.5 are observed in such models.

 
 Recent global non-ideal MHD simulations of magnetised disks, including more comprehensive disk microphysics and evolving both hemispheres, have confirmed these early results and shown that pronounced asymmetries between the lower and upper disk halves of the disk can develop and launch asymmetric outflows which persist over timescales of at least 1000 times the orbital period at the inner disk radius \citep{Bethune2017, Bai2017, Gressel2020, Riols2020}. However, most of these recent simulations do not currently include the very central regions of the disk (r~$\leq$~1~au) from which the high-velocity jets are launched.
 

\subsection{The Origin of Jet Axis Wiggling}
\label{sec:discuss_jetwiggle}

As seen in Section~\ref{sec:discuss_footpoint}, the DO~Tau bipolar jet shows a quasi sinusoidal small amplitude wiggling of its axis. Similar wiggling has been recently revealed at the base of other T Tauri jets \citep[e.g.][]{Dougados2000, Anglada2007, Garufi2019}. This wiggling is often interpreted as tracing the dynamical perturbation due to an unseen companion.
A companion can induce wiggling of the jet axis in two different ways. The first one is orbital motion of the jet source (\citealp{Masciadri2002}; hereafter `orbital scenario'). Alternatively, a mis-aligned companion can cause retrograde precession of the disk rotation axis around the orbital axis \cite{Papaloizou1995}, and hence of the associated jet axis (\citep{Terquem1999}; hereafter `precession scenario'). \citet{SheikhnezamiFendt2018} have studied the onset of jet/disk precession in a binary system and showed that moderate mis-alignments between the binary and disk orbital planes ($< 10^{\circ}$) are required for persistent MHD jet launching. The small wiggling angle detected here would therefore be consistent with such scenario. We discuss below the two possible binary wiggling mechanisms, and show that both scenarii met difficulties in reproducing the pattern of the DO Tau bipolar jet wiggle. We then discuss an alternative jet precession mechanism related to a possible warping instability in disks launching magnetically-driven outflows, proposed by \citet{Lai2003}.

\subsubsection{Wiggling due to binarity}
\label{sec:binary scenario}

In principle, we can distinguish between the orbital and precession motion induced by a companion by examining the symmetry pattern of the wiggling between the jet and counter-jet: orbital motion produces a mirror-symmetry, while jet precession produces an S-shaped symmetry \citep{Masciadri2002}. For DO Tau, however, the effect is complicated by the velocity asymmetry between the two jet lobes, and by the limited field of view of our NIFS observations. To examine these two scenarii, we use the formalism and equations first introduced by \citet{Masciadri2002}, and further developed in \citet{Anglada2007} and \citet{Estalella2012}, to model the wiggling in the HH~30 bipolar jet.  


\paragraph{i) {\it Orbital scenario}}
\label{sec:orbital scenario}

\begin{figure*}
    \centering
    \includegraphics[width=0.7\textwidth]{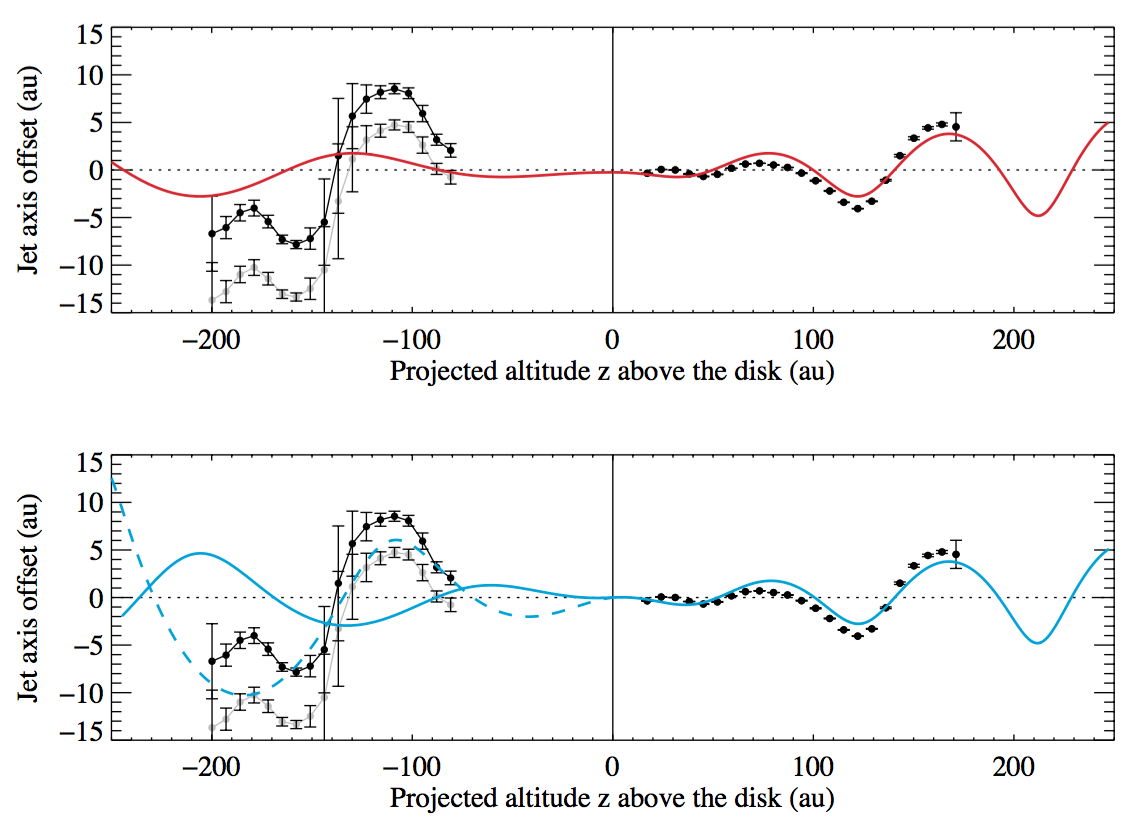}
    \caption{Wiggle in the jet axis of DO Tau: Black dots with associated error bars show the jet axis displacement, measured with respect to the blue-shifted jet PA. On the red-shifted jet side (left), grey dots show measurements before correction for a possible misalignment of 1$^{\circ}$ of the red-shifted jet. {\bf Top panel:} The red solid curve shows the orbital solution fitted to the blue-shifted jet axis wiggling. {\bf Bottom panel:} The solid cyan curve shows the precession solution fitted to the blue-shifted jet axis wiggling. The dashed cyan curve shows the same precession solution computed for the red-shifted jet, but adjusted for a phase shift and an increased precession angle (by a factor 2.5). See text for more details.} 
    \label{fig:wiggling} 
\end{figure*}

In the orbital scenario, the launching point of the jet moves as the jet source orbits the center of mass of the system (where the jet source can be either the primary or secondary component). We further assume that the jet axis and the binary orbital axis coincide, i.e. the binary orbital plane and the disk plane of the jet source are aligned.

Following \citet{Anglada2007}, we define m$_1$ as the mass of the jet source and $\mu=\text{m}_2/\text{m}_{\text{tot}}$ as the ratio between the mass of the companion and the total mass of the system. Let $a$ be the binary separation and r$_0$ the orbital radius of the jet source with respect to the center of mass of the system ($\text{r}_0 = \mu a$). 
Using the wiggling parameters on the blue-shifted side  with equations (6) \& (7) from \citet{Anglada2007}, and including an uncertainty of $5^{\circ}$ for the jet inclination and 5~\kms for the blue-shifted jet line of sight velocity, we derive for the jet source an orbital velocity of v$_0$=1. $\pm$ 0.45 \kms and an orbital radius of r$_0$=0.32 $\pm$ 0.13~au. The orbital period is then $\tau_{\rm orb}$=9.5 $\pm$ 5.6~years. This best-fit solution is represented in Figure~\ref{fig:wiggling}, top panel. This solution reproduces the amplitude and spatial wavelength of the jet axis wiggling on the blue-shifted side but fails to reproduce the amplitude of the wiggles on the red-shifted side.

Indeed, the larger amplitude wiggling on the fastest jet side is in contradiction with the expectation from the orbital model. In the orbital scenario, the amplitude of the wiggles is set by the ratio between $V_0$, the orbital velocity of the jet source around the center of mass of the system, and $V_{jet}$. Given that the receding jet has roughly twice the velocity of the approaching jet, we expect the opening angle of the wiggles to be two times smaller on the red-shifted side, which is clearly not observed.

This orbital solution also predicts a periodic variation in line-of-sight velocity of amplitude $\text{v}_0 \times \sin{\text{i}}$ = 0.2-0.7 \kms, which is an order of magnitude smaller than the observed radial velocity variations along the blue-shifted jet ($\pm 5$ \kms over the inner 200~au) and within our estimated uncertainties on the centroid velocities. So such a variation could be hidden in larger amplitude variations due to, for example, time variability in the ejection velocity.



\paragraph{ii) {\it Precession scenario}}

Wiggling of the jet axis could also be the signature of precession driven by tidal interactions between the disk and a companion in a non-coplanar orbit. This scenario has been investigated in the context of the HH~30 bipolar jet by \citet{Anglada2007,Estalella2012}. 
We assume that the jet axis precesses with a precession angle $\beta$ and precession period $\tau_{\text{pre}}$ and that the jet precession axis coincides with the average jet PA. We show in Figure~\ref{fig:wiggling} (solid blue curve) the {\sl equivalent} precession solution obtained by setting the precession period equal to the period of the orbital solution ($\tau_{\text{pre}}=9.5 \pm 5.6$~yrs) and the precession angle $\beta$ such that $\tan(\beta)$ = $\tan(\theta_{\text{obs}}) \times \sin{\text{i}}$, which gives $\beta$= 0.52 $\pm$ 0.22 $\degr$ for the approaching jet. 

Similar to the orbital solution, the spatial wavelength of the wiggles is expected to be twice as large on the receding side as on the approaching side. However,contrary to the orbital scenario, the apparent opening angle of the wiggles does not depend on the jet velocity since it is set by the precession angle. Such a solution also predicts a periodic variation of the line of sight velocity of amplitude $\pm v_j \sin(\beta) \sin(i_{\text{jet}}) \simeq 0.4$ \kms, very similar to the orbital case. These variations are much smaller than both our estimated uncertainties and the observed radial velocity variations along the jet.

Again, the agreement is not good on the receding jet side. A better match can be obtained by introducing a phase shift between the precession of the red- and blue-shifted jets and increasing the precession angle by a factor $\simeq$ 2-2.5 for the red-shifted jet (dashed blue curve in Fig.\ref{fig:wiggling}). Such a solution would  be possible if the red- and blue-shifted jets originate from annuli at different radii in a warped disk (each disk annulus would make a different angle in 3D with respect to the orbital plane). 


In the case of a companion in an inclined orbit perturbing the circumstellar disk of the jet source, a general expression for the precessional period was originally derived by \citet{Papaloizou1995} and \citet{Terquem1998} for arbitrary disk surface density distributions and rotation curves, assuming rigid body precession and a circular orbit. The orbital period of the companion is found to be significantly smaller than the disk precession period, by at least one order of magnitude if $\mu < 1$ (see Equation 24 in \citet{Terquem1998}). 
In the DO Tau case, however, this would imply orbital periods shorter than one year, which would be inconsistent with an orbit 2-4 times wider than DO Tau's disk (which has $R_d$ > 150 au, see Section~3.1).

An alternative scenario is a planetary mass companion in a misaligned orbit within the primary disk itself. Recent works have studied the impact of such a companion on the disk \citep{Xiang-Gruess2013,Nealson2018,Zhu2019}.
For sufficiently massive planets, a gap is created at the orbital radius which can divide the disk in two parts. The inner disk then precesses at a much faster rate than the outer disk. The relationship between the companion orbital period $\tau_\text{orb}$ and the inner disk precession period $\tau_\text{pre}$ is then  given by a similar expression than in the distant binary scenario (see Eq. 27 in \citet{Zhu2019}):
\begin{equation}
\frac{\tau_\text{orb}}{\tau_\text{pre}} = \frac{3}{8} \text{cos}(\text{i}_\text{p}) \left( \frac{\mu}{\sqrt(1-\mu)} \right) \sigma^{3/2} 
\end{equation}
where $\text{i}_\text{p}$ is the angle between the disk and the companion orbital angular momentum vectors and $\sigma=\text{R}_\text{d}/\text{a}$ the ratio of the radius of the inner disk to the companion separation. In the mis-aligned planetary mass companion scenario $\sigma \simeq 1$.
While this ratio is 5 times larger than in the previous "distant binary" scenario (where $\sigma \simeq$ 1/3), it is still much smaller than 1 for $\mu << 1$ (planetary mass companion).
In the case of DO~Tau, with $\text{i}_\text{p}  \simeq \beta = 0.5 \degr$, we thus have:
\begin{equation}
\frac{\tau_{\rm orb}}{\tau_{\rm pre}} \simeq 0.37 \frac{\mu}{\sqrt(1-\mu)} \simeq 0.37 \mu~ \text{for}~ \mu << 1
\end{equation}

Additional constraints can be obtained by requiring that the angular effect produced by the orbital motion of the jet source is negligible compared to the precession angle, i.e. $\frac{\text{V}_0}{\text{V}_\text{j}} \le \text{tan}(\beta)$, which translates here into v$_0 \le 1.4$ \kms (taking into account uncertainties on $\text{V}_\text{j}$ and $\beta$).
The total mass of the system $\text{m}_{\text{tot}}$ and orbital velocity $\text{V}_0$ can be expressed as a function of the orbital parameters by:
\begin{equation}
\begin{split}
\left( \frac{\text{m}_{\rm tot}}{\text{M}_{\odot}} \right) =  \mu^{-3}  \left( \frac{\text{r}_0}{\rm au} \right)^3  \left( \frac{\tau_\text{orb}}{\rm yr} \right) ^{-2} 
\end{split}
\end{equation}

\begin{equation}
\begin{split}
{V}_0 = 2 \pi \frac{\text{r}_0}{\tau_\text{orb}} 
\end{split}
\end{equation}

Combining these equations, the requirements that $\text{v}_0 \le 1.4$ \kms and $ \text{m}_{\text{tot}}\simeq 0.6$ M$_{\odot}$ then imply $\mu \le 0.02$, ie. companion masses $\le$ 12 M$_{Jup}$, and separations $\text{a} \leq$ 0.15~au. 

Therefore the precession scenario requires a massive planetary mass companion in a very close orbit (separation less than 0.15 au) with a small mis-alignment of its orbit (a few degrees). It is not clear how such small  misalignments could be preserved over long timescales. \cite{Zhu2019,Xiang-Gruess2013} show that, in standard disk conditions, a massive enough planet ($>$ 6 $M_{J}$) is required for the inclination damping timescale to exceed the precession timescale at small misalignments.
With this additional constraint, the precession solution requires a companion mass in the range 6-12~M$_{Jup}$ with separations 0.1-0.15~au. 
Such a solution would still allow the launching of the jet but seems unlikely as the survival of the inner disk may be affected at such small separations. It is also not clear how the red- and blue-shifted jets would precess with slightly different solutions since this scenario predicts solid body precession of the inner disk.



\subsubsection{Precession induced by magnetic torques}


Given that the binary models face significant issues in reproducing the observed wiggling in the DO Tau bipolar jet, we investigate an alternative mechanism. 
\citet{Lai2003} recently proposed that magnetic torques associated with the outflow may cause warping instability of the accretion disk, and hence precession of the disk rotation axis and associated jet axis. This study finds that the growth timescale for the disk warp, and so the precession period, is of the order of the radial accretion time of the disk, r/|v$_r$|. The study suggests that a warped, precessing disk-outflow system may be an alternative (and preferred) state for an accretion disk threaded by large-scale magnetic fields. 

Here, we apply this idea to the case of DO Tau. We find the disk radial accretion timescale, r/|v$_r$|, by taking the radial distance travelled, r, as the outer limit on our jet launching radius, r$_0$ = 0.5~au $\pm$ 0.1~au. In a standard viscous $\alpha$-disk model \citep{Shakura1973}, the radial accretion velocity is given by v$_r$ $\sim$ $\alpha$ (h/r) $c_s$ under the assumption of a thin disk (i.e. h/r $<<$ 1). Typically, $\alpha = 10^{-2}-10^{-3}$ and h/r$=0.1$, giving radial velocities which are very subsonic ($\simeq 10^{-3} c_s$). However, in a disk launching a jet through magneto-centrifugal processes, accretion radial velocities are significantly increased with respect to the standard viscous case, due to the efficient extraction of angular momentum provided by the magnetised wind. 
In such as case, radial velocities are typically expected to fall between 0.1 and 1 c$_s$ 
\citep[see e.g.][]{Combet2008}. Taking a gas temperature of 1000 K, which gives a sound speed $c_s$ $\sim$ 3 \kms, we obtain a typical radial accretion timescale ranging from $0.7$ to $7$~yrs. This rough estimate is marginally compatible with the derived jet precession period of 10 $\pm$ 2.7 years, suggesting that such a mechanism could be at work in the DO~Tau disk. This disk warp could also account for a difference in PA of 1$^{\degr}$ between the blue- and red-shifted jets, as well as for a slightly different precession solution for the two jets if they arise from annuli at different radii in the disk.   

\section{Conclusions}\label{sec:conclusions}

We present a case study of the DO Tau bipolar jet, examining the jet base with high resolution spectro-imaging, in order to constrain jet launching models. The main conclusions are the following:

- The DO Tau blue-shifted jet is among the narrowest and initially most collimated Class II jets identified so far. Strong collimation (with semi-opening angle $\leq$ 1.5$^{\circ}$) is achieved within deprojected distance from the source $z_0$=40~au. The derived opening angle is twice smaller than the estimated Mach opening angle.
The jet base is very narrow, with a jet radius $<$ 4~au at $z_0$=40~au. This is similar to recent observations at the base of the Class 0 HH\,212 SiO jet \citep{Lee2017}. These results strongly support a collimation mechanism independent of evolutionary stage, as already suggested by \citet{Cabrit2007}, which in turn strongly supports a magnetic collimation mechanism. \\
- The DO Tau red-shifted jet is not detected until projected distances from the source z $\ge$ 0.5$^{\prime\prime}$, consistent with occultation by the gaseous disk observed with ALMA. It then shows a similar collimation to its blue-shifted counter-part.\\ 
- For the DO Tau blue-shifted jet, terminal velocities are reached within $z_0$=40~au and show only small variability (less than 20\%) thereafter on a time-span of 15 years.  This suggests weak shocks, if present at all. Indeed, no emission knots can be clearly identified in the blue-shifted jet [\ion{Fe}{II}] emission.\\
- We measure a blue-shifted jet electron number density of $5 \times 10^4$~cm$^{-3}$ close to the source (z$\sim$0$\farcs$3) followed by a sharp decline and plateau to $1 \times 10^4$~cm$^{-3}$ at z $\sim$0$\farcs$4-1$\arcsec$ from the source. Combined with our velocity measurements, we derive a jet mass flux of $\dot{M}_{jet}$ $\sim$ 5 $\times$ 10$^{-9}$ M$_{\odot}$ yr$^{-1}$ for the DO Tau blue-shifted jet. Assuming a similar mass flux for the red-shifted jet yields an overall ejection to accretion mass flux ratio of $\dot{M}_{jet}$ $\times$2/$\dot{M}_{acc}$ $\sim$ 0.07. \\
- No systematic difference in peak centroid velocity is observed across the blue-shifted jet. Our observations put a 3~$\sigma$ upper limit on jet rotation velocities of V$_{\phi}$ $\leq$ 6.3 and 8.7~km~s$^{-1}$ for the blue and red-shifted jets, respectively, at jet radii of r=14~au. This yields specific angular momentum of $r \times V_{\rm phi} \leq$ 87 and 122 ~au~km~s$^{-1}$ for the blue and red-shifted jets, respectively. Under a \emph{steady} and \emph{axisymmetric} MHD disk-wind launching process assumption, this implies a jet launch radius in the disk plane of \textbf{r$_{0}$} $\leq$ 0.5~au and \textbf{r$_{0}$} $\leq$ 0.3~au for the blue and red-shifted jets, respectively. We discuss the possibility that such difference derives from an asymmetric launch mechanism in the two disk hemispheres. An origin so close to the star of the [\ion{Fe}{II}] jet is consistent with the absence of significant Fe gas phase depletion. However, we note that the detection of wiggling in the jet may call into question the assumption of axisymmetry, under which these values have been determined.\\ 
- We confirm the previously reported striking velocity asymmetry between jet and counter-jet, of a factor 1.7. This asymmetry is established close to the source and stays constant over a time span of at least 7 years corresponding to $\simeq$ 500 orbital timescales at the inner disk region. This long lasting asymmetry is in accordance with predictions from recent simulations of magnetised disk-winds. \\
- We identify jet axis wiggling in both lobes with an apparent semi-opening angle of 1.3 $\pm$ 0.5$^{\circ}$ and spatial wavelength of 90 $\pm$ 5~au on the blue-shifted side. The spatial wavelength of the wiggles is poorly constrained on the red-shifted side but the opening angle appears larger at 2.6-4.7$^{\circ}$.
The larger amplitude wiggling observed for the faster (red-shifted) jet excludes orbital motion of the jet source as the origin of the wiggling. Precession of the jet axis can be fitted to the observed wiggling on both sides if different precession angles and precession phases are assumed. Such a solution would require the blue- and red-shifted jets to originate from rings at different radii in the disk, and to precess slightly differently.
A warping instability, induced by the launching of magnetic disk-wind, is a promising mechanism to explain the disk precession and wiggling in the DO Tau bipolar jet. \\

Overall, the observed properties of the DO Tau bipolar jet are consistent with properties of a disk-wind model for jet launching, provided it originates in the inner regions of the disk. Our study shows the power of high resolution spectro-imaging observations close to jet base in imposing constraints on competing jet launching models. However, further case studies are required to build statistics on {\it asymmetric} bipolar jets, if jet launching is to be fully understood with all its consequent implications for magnetic field strengths in the inner regions of protoplanetary disks. 


\begin{acknowledgements}
This work has benefited from the ULYSSES PHC France Ireland cooperative program. This work was supported by the "Programme National de Physique Stellaire" (PNPS) of CNRS/INSU co-funded by CEA and CNES. JE acknowledges funding from the UCD Physics Scholarship in Research and Teaching. FB acknowledges support from the project PRIN-INAF-MAIN-STREAM 2017 “Protoplanetary disks seen through the eyes of new-generation instruments”. Based on observations obtained at the Gemini Observatory, which is operated by the Association of Universities for Research in Astronomy, Inc., under a cooperative agreement with the NSF on behalf of the Gemini partnership: the National Science Foundation (United States), National Research Council (Canada), CONICYT (Chile), Ministerio de Ciencia, Tecnolog\'{i}a e Innovaci\'{o}n Productiva (Argentina), Minist\'{e}rio da Ci\^{e}ncia, Tecnologia e Inova\c{c}\~{a}o (Brazil), and Korea Astronomy and Space Science Institute (Republic of Korea).  
\end{acknowledgements}


\bibliography{bibliography.bib}

\begin{thebibliography}{99}
\expandafter\ifx\csname natexlab\endcsname\relax\def\natexlab#1{#1}\fi

\bibitem[{{Agra-Amboage} {et~al.}(2014){Agra-Amboage}, {Cabrit}, {Dougados},
  {Kristensen}, {Ibgui}, \& {Reunanen}}]{Agra-Amboage2014}
{Agra-Amboage}, V., {Cabrit}, S., {Dougados}, C., {et~al.} 2014, \aap, 564, A11

\bibitem[{{Agra-Amboage} {et~al.}(2011){Agra-Amboage}, {Dougados}, {Cabrit}, \&
  {Reunanen}}]{Agra-Amboage2011}
{Agra-Amboage}, V., {Dougados}, C., {Cabrit}, S., \& {Reunanen}, J. 2011, \aap,
  532, A59

\bibitem[{{Anderson} {et~al.}(2003){Anderson}, {Li}, {Krasnopolsky}, \&
  {Blandford}}]{Anderson2003}
{Anderson}, J.~M., {Li}, Z.-Y., {Krasnopolsky}, R., \& {Blandford}, R.~D. 2003,
  \apjl, 590, L107

\bibitem[{{Anglada} {et~al.}(2007){Anglada}, {L{\'o}pez}, {Estalella},
  {Masegosa}, {Riera}, \& {Raga}}]{Anglada2007}
{Anglada}, G., {L{\'o}pez}, R., {Estalella}, R., {et~al.} 2007, \aj, 133, 2799

\bibitem[{{Antoniucci} {et~al.}(2016){Antoniucci}, {Podio}, {Nisini},
  {Bacciotti}, {Lagadec}, {Sissa}, {La Camera}, {Giannini}, {Schmid},
  {Gratton}, {Turatto}, {Desidera}, {Bonnefoy}, {Chauvin}, {Dougados},
  {Bazzon}, {Thalmann}, \& {Langlois}}]{Antoniucci2016}
{Antoniucci}, S., {Podio}, L., {Nisini}, B., {et~al.} 2016, \aap, 593, L13

\bibitem[{{Bacciotti} \& {Eisl{\"o}ffel}(1999)}]{Bacciotti1999}
{Bacciotti}, F. \& {Eisl{\"o}ffel}, J. 1999, \aap, 342, 717

\bibitem[{{Bacciotti} {et~al.}(2018){Bacciotti}, {Girart}, {Padovani}, {Podio},
  {Paladino}, {Testi}, {Bianchi}, {Galli}, {Codella}, {Coffey}, {Favre}, \&
  {Fedele}}]{Bacciotti2018}
{Bacciotti}, F., {Girart}, J.~M., {Padovani}, M., {et~al.} 2018, \apjl, 865,
  L12

\bibitem[{{Bacciotti} {et~al.}(2000){Bacciotti}, {Mundt}, {Ray},
  {Eisl{\"o}ffel}, {Solf}, \& {Camezind}}]{Bacciotti2000}
{Bacciotti}, F., {Mundt}, R., {Ray}, T.~P., {et~al.} 2000, \apjl, 537, L49

\bibitem[{{Bacciotti} {et~al.}(2002){Bacciotti}, {Ray}, {Mundt},
  {Eisl{\"o}ffel}, \& {Solf}}]{Bacciotti2002}
{Bacciotti}, F., {Ray}, T.~P., {Mundt}, R., {Eisl{\"o}ffel}, J., \& {Solf}, J.
  2002, \apj, 576, 222

\bibitem[{{Bai}(2017)}]{Bai2017}
{Bai}, X.-N. 2017, \apj, 845, 75

\bibitem[{{Baruteau} {et~al.}(2014){Baruteau}, {Crida}, {Paardekooper},
  {Masset}, {Guilet}, {Bitsch}, {Nelson}, {Kley}, \&
  {Papaloizou}}]{Baruteau2014}
{Baruteau}, C., {Crida}, A., {Paardekooper}, S.~J., {et~al.} 2014, in
  Protostars and Planets VI, ed. H.~{Beuther}, R.~S. {Klessen}, C.~P.
  {Dullemond}, \& T.~{Henning}, 667

\bibitem[{{Bergner} {et~al.}(2019){Bergner}, {{\"O}berg}, {Bergin}, {Loomis},
  {Pegues}, \& {Qi}}]{Bergner2019}
{Bergner}, J.~B., {{\"O}berg}, K.~I., {Bergin}, E.~A., {et~al.} 2019, \apj,
  876, 25

\bibitem[{{B{\'e}thune} {et~al.}(2017){B{\'e}thune}, {Lesur}, \&
  {Ferreira}}]{Bethune2017}
{B{\'e}thune}, W., {Lesur}, G., \& {Ferreira}, J. 2017, \aap, 600, A75

\bibitem[{{Blandford} \& {Payne}(1982)}]{BlandfordPayne1982}
{Blandford}, R.~D. \& {Payne}, D.~G. 1982, \mnras, 199, 883

\bibitem[{{Cabrit} {et~al.}(2007){Cabrit}, {Codella}, {Gueth}, {Nisini},
  {Gusdorf}, {Dougados}, \& {Bacciotti}}]{Cabrit2007}
{Cabrit}, S., {Codella}, C., {Gueth}, F., {et~al.} 2007, \aap, 468, L29

\bibitem[{{Casse} \& {Ferreira}(2000)}]{Casse2000}
{Casse}, F. \& {Ferreira}, J. 2000, \aap, 361, 1178

\bibitem[{{Casse} \& {Keppens}(2002)}]{CasseKeppens2002}
{Casse}, F. \& {Keppens}, R. 2002, \apj, 581, 988

\bibitem[{{Casse} \& {Keppens}(2004)}]{CasseKeppens2004}
{Casse}, F. \& {Keppens}, R. 2004, \apj, 601, 90

\bibitem[{{Cerqueira} {et~al.}(2006){Cerqueira}, {Vel{\'a}zquez}, {Raga},
  {Vasconcelos}, \& {de Colle}}]{Cerqueira2006}
{Cerqueira}, A.~H., {Vel{\'a}zquez}, P.~F., {Raga}, A.~C., {Vasconcelos},
  M.~J., \& {de Colle}, F. 2006, \aap, 448, 231

\bibitem[{{Chrysostomou} {et~al.}(2008){Chrysostomou}, {Bacciotti}, {Nisini},
  {Ray}, {Eisl{\"o}ffel}, {Davis}, \& {Takami}}]{Chrysostomou2008}
{Chrysostomou}, A., {Bacciotti}, F., {Nisini}, B., {et~al.} 2008, \aap, 482,
  575

\bibitem[{{Coffey}(2017)}]{Coffey2017}
{Coffey}, D. 2017, Nature Astronomy, 1, 0180

\bibitem[{{Coffey} {et~al.}(2011){Coffey}, {Bacciotti}, {Chrysostomou},
  {Nisini}, \& {Davis}}]{Coffey2011}
{Coffey}, D., {Bacciotti}, F., {Chrysostomou}, A., {Nisini}, B., \& {Davis}, C.
  2011, \aap, 526, A40

\bibitem[{{Coffey} {et~al.}(2008){Coffey}, {Bacciotti}, \&
  {Podio}}]{Coffey2008}
{Coffey}, D., {Bacciotti}, F., \& {Podio}, L. 2008, \apj, 689, 1112

\bibitem[{{Coffey} {et~al.}(2004){Coffey}, {Bacciotti}, {Woitas}, {Ray}, \&
  {Eisl{\"o}ffel}}]{Coffey2004}
{Coffey}, D., {Bacciotti}, F., {Woitas}, J., {Ray}, T.~P., \& {Eisl{\"o}ffel},
  J. 2004, \apj, 604, 758

\bibitem[{{Coffey} {et~al.}(2015){Coffey}, {Dougados}, {Cabrit}, {Pety}, \&
  {Bacciotti}}]{Coffey2015}
{Coffey}, D., {Dougados}, C., {Cabrit}, S., {Pety}, J., \& {Bacciotti}, F.
  2015, \apj, 804, 2

\bibitem[{{Combet} \& {Ferreira}(2008)}]{Combet2008}
{Combet}, C. \& {Ferreira}, J. 2008, \aap, 479, 481

\bibitem[{{Dougados} {et~al.}(2000){Dougados}, {Cabrit}, {Lavalley}, \&
  {M{\'e}nard}}]{Dougados2000}
{Dougados}, C., {Cabrit}, S., {Lavalley}, C., \& {M{\'e}nard}, F. 2000, \aap,
  357, L61

\bibitem[{{Dyda} {et~al.}(2015){Dyda}, {Lovelace}, {Ustyugova}, {Lii},
  {Romanova}, \& {Koldoba}}]{Dyda2015}
{Dyda}, S., {Lovelace}, R.~V.~E., {Ustyugova}, G.~V., {et~al.} 2015, \mnras,
  450, 481

\bibitem[{{Edwards} {et~al.}(1987){Edwards}, {Cabrit}, {Strom}, {Heyer},
  {Strom}, \& {Anderson}}]{Edwards1987}
{Edwards}, S., {Cabrit}, S., {Strom}, S.~E., {et~al.} 1987, \apj, 321, 473

\bibitem[{{Eisner} {et~al.}(2014){Eisner}, {Hillenbrand}, \&
  {Stone}}]{Eisner2014}
{Eisner}, J.~A., {Hillenbrand}, L.~A., \& {Stone}, J.~M. 2014, \mnras, 443,
  1916

\bibitem[{{Estalella} {et~al.}(2012){Estalella}, {L{\'o}pez}, {Anglada},
  {G{\'o}mez}, {Riera}, \& {Carrasco-Gonz{\'a}lez}}]{Estalella2012}
{Estalella}, R., {L{\'o}pez}, R., {Anglada}, G., {et~al.} 2012, \aj, 144, 61

\bibitem[{{Fendt}(2011)}]{Fendt2011}
{Fendt}, C. 2011, \apj, 737, 43

\bibitem[{{Fendt} \& {Sheikhnezami}(2013)}]{Fendt2013}
{Fendt}, C. \& {Sheikhnezami}, S. 2013, \apj, 774, 12

\bibitem[{{Fendt} \& {Zinnecker}(1998)}]{FendtZinnecker1998}
{Fendt}, C. \& {Zinnecker}, H. 1998, \aap, 334, 750

\bibitem[{{Fern{\'a}ndez-L{\'o}pez} {et~al.}(2020){Fern{\'a}ndez-L{\'o}pez},
  {Zapata}, {Rodr{\'\i}guez}, {Vazzano}, {Guzm{\'a}n}, \&
  {L{\'o}pez}}]{FernandezLopez2020}
{Fern{\'a}ndez-L{\'o}pez}, M., {Zapata}, L.~A., {Rodr{\'\i}guez}, L.~F.,
  {et~al.} 2020, \aj, 159, 171

\bibitem[{{Ferreira}(1997)}]{Ferreira1997}
{Ferreira}, J. 1997, \aap, 319, 340

\bibitem[{{Ferreira} {et~al.}(2006){Ferreira}, {Dougados}, \&
  {Cabrit}}]{Ferreira2006}
{Ferreira}, J., {Dougados}, C., \& {Cabrit}, S. 2006, \aap, 453, 785

\bibitem[{{Frank} {et~al.}(2014){Frank}, {Ray}, {Cabrit}, {Hartigan}, {Arce},
  {Bacciotti}, {Bally}, {Benisty}, {Eisl{\"o}ffel}, {G{\"u}del}, {Lebedev},
  {Nisini}, \& {Raga}}]{Frank2014}
{Frank}, A., {Ray}, T.~P., {Cabrit}, S., {et~al.} 2014, in Protostars and
  Planets VI, ed. H.~{Beuther}, R.~S. {Klessen}, C.~P. {Dullemond}, \&
  T.~{Henning}, 451

\bibitem[{{Gaia Collaboration} {et~al.}(2018){Gaia Collaboration}, {Brown},
  {Vallenari}, {Prusti}, {de Bruijne}, {Babusiaux}, {Bailer-Jones}, {Biermann},
  {Evans}, {Eyer}, {Jansen}, {Jordi}, {Klioner}, {Lammers}, {Lindegren},
  {Luri}, {Mignard}, {Panem}, {Pourbaix}, {Randich}, {Sartoretti}, {Siddiqui},
  {Soubiran}, {van Leeuwen}, {Walton}, {Arenou}, {Bastian}, {Cropper},
  {Drimmel}, {Katz}, {Lattanzi}, {Bakker}, {Cacciari}, {Casta{\~n}eda},
  {Chaoul}, {Cheek}, {De Angeli}, {Fabricius}, {Guerra}, {Holl}, {Masana},
  {Messineo}, {Mowlavi}, {Nienartowicz}, {Panuzzo}, {Portell}, {Riello},
  {Seabroke}, {Tanga}, {Th{\'e}venin}, {Gracia-Abril}, {Comoretto},
  {Garcia-Reinaldos}, {Teyssier}, {Altmann}, {Andrae}, {Audard},
  {Bellas-Velidis}, {Benson}, {Berthier}, {Blomme}, {Burgess}, {Busso},
  {Carry}, {Cellino}, {Clementini}, {Clotet}, {Creevey}, {Davidson}, {De
  Ridder}, {Delchambre}, {Dell'Oro}, {Ducourant},
  {Fern{\'a}ndez-Hern{\'a}ndez}, {Fouesneau}, {Fr{\'e}mat}, {Galluccio},
  {Garc{\'\i}a-Torres}, {Gonz{\'a}lez-N{\'u}{\~n}ez}, {Gonz{\'a}lez-Vidal},
  {Gosset}, {Guy}, {Halbwachs}, {Hambly}, {Harrison}, {Hern{\'a}ndez},
  {Hestroffer}, {Hodgkin}, {Hutton}, {Jasniewicz}, {Jean-Antoine-Piccolo},
  {Jordan}, {Korn}, {Krone-Martins}, {Lanzafame}, {Lebzelter}, {L{\"o}ffler},
  {Manteiga}, {Marrese}, {Mart{\'\i}n-Fleitas}, {Moitinho}, {Mora}, {Muinonen},
  {Osinde}, {Pancino}, {Pauwels}, {Petit}, {Recio-Blanco}, {Richards},
  {Rimoldini}, {Robin}, {Sarro}, {Siopis}, {Smith}, {Sozzetti}, {S{\"u}veges},
  {Torra}, {van Reeven}, {Abbas}, {Abreu Aramburu}, {Accart}, {Aerts},
  {Altavilla}, {{\'A}lvarez}, {Alvarez}, {Alves}, {Anderson}, {Andrei},
  {Anglada Varela}, {Antiche}, {Antoja}, {Arcay}, {Astraatmadja}, {Bach},
  {Baker}, {Balaguer-N{\'u}{\~n}ez}, {Balm}, {Barache}, {Barata}, {Barbato},
  {Barblan}, {Barklem}, {Barrado}, {Barros}, {Barstow}, {Bartholom{\'e}
  Mu{\~n}oz}, {Bassilana}, {Becciani}, {Bellazzini}, {Berihuete}, {Bertone},
  {Bianchi}, {Bienaym{\'e}}, {Blanco-Cuaresma}, {Boch}, {Boeche}, {Bombrun},
  {Borrachero}, {Bossini}, {Bouquillon}, {Bourda}, {Bragaglia}, {Bramante},
  {Breddels}, {Bressan}, {Brouillet}, {Br{\"u}semeister}, {Brugaletta},
  {Bucciarelli}, {Burlacu}, {Busonero}, {Butkevich}, {Buzzi}, {Caffau},
  {Cancelliere}, {Cannizzaro}, {Cantat-Gaudin}, {Carballo}, {Carlucci},
  {Carrasco}, {Casamiquela}, {Castellani}, {Castro-Ginard}, {Charlot},
  {Chemin}, {Chiavassa}, {Cocozza}, {Costigan}, {Cowell}, {Crifo}, {Crosta},
  {Crowley}, {Cuypers}, {Dafonte}, {Damerdji}, {Dapergolas}, {David}, {David},
  {de Laverny}, {De Luise}, {De March}, {de Martino}, {de Souza}, {de Torres},
  {Debosscher}, {del Pozo}, {Delbo}, {Delgado}, {Delgado}, {Di Matteo},
  {Diakite}, {Diener}, {Distefano}, {Dolding}, {Drazinos}, {Dur{\'a}n},
  {Edvardsson}, {Enke}, {Eriksson}, {Esquej}, {Eynard Bontemps}, {Fabre},
  {Fabrizio}, {Faigler}, {Falc{\~a}o}, {Farr{\`a}s Casas}, {Federici},
  {Fedorets}, {Fernique}, {Figueras}, {Filippi}, {Findeisen}, {Fonti},
  {Fraile}, {Fraser}, {Fr{\'e}zouls}, {Gai}, {Galleti}, {Garabato},
  {Garc{\'\i}a-Sedano}, {Garofalo}, {Garralda}, {Gavel}, {Gavras}, {Gerssen},
  {Geyer}, {Giacobbe}, {Gilmore}, {Girona}, {Giuffrida}, {Glass}, {Gomes},
  {Granvik}, {Gueguen}, {Guerrier}, {Guiraud}, {Guti{\'e}rrez-S{\'a}nchez},
  {Haigron}, {Hatzidimitriou}, {Hauser}, {Haywood}, {Heiter}, {Helmi}, {Heu},
  {Hilger}, {Hobbs}, {Hofmann}, {Holland}, {Huckle}, {Hypki}, {Icardi},
  {Jan{\ss}en}, {Jevardat de Fombelle}, {Jonker}, {Juh{\'a}sz}, {Julbe},
  {Karampelas}, {Kewley}, {Klar}, {Kochoska}, {Kohley}, {Kolenberg},
  {Kontizas}, {Kontizas}, {Koposov}, {Kordopatis}, {Kostrzewa-Rutkowska},
  {Koubsky}, {Lambert}, {Lanza}, {Lasne}, {Lavigne}, {Le Fustec}, {Le
  Poncin-Lafitte}, {Lebreton}, {Leccia}, {Leclerc}, {Lecoeur-Taibi},
  {Lenhardt}, {Leroux}, {Liao}, {Licata}, {Lindstr{\o}m}, {Lister}, {Livanou},
  {Lobel}, {L{\'o}pez}, {Managau}, {Mann}, {Mantelet}, {Marchal}, {Marchant},
  {Marconi}, {Marinoni}, {Marschalk{\'o}}, {Marshall}, {Martino}, {Marton},
  {Mary}, {Massari}, {Matijevi{\v{c}}}, {Mazeh}, {McMillan}, {Messina},
  {Michalik}, {Millar}, {Molina}, {Molinaro}, {Moln{\'a}r}, {Montegriffo},
  {Mor}, {Morbidelli}, {Morel}, {Morris}, {Mulone}, {Muraveva}, {Musella},
  {Nelemans}, {Nicastro}, {Noval}, {O'Mullane}, {Ord{\'e}novic},
  {Ord{\'o}{\~n}ez-Blanco}, {Osborne}, {Pagani}, {Pagano}, {Pailler},
  {Palacin}, {Palaversa}, {Panahi}, {Pawlak}, {Piersimoni}, {Pineau}, {Plachy},
  {Plum}, {Poggio}, {Poujoulet}, {Pr{\v{s}}a}, {Pulone}, {Racero}, {Ragaini},
  {Rambaux}, {Ramos-Lerate}, {Regibo}, {Reyl{\'e}}, {Riclet}, {Ripepi}, {Riva},
  {Rivard}, {Rixon}, {Roegiers}, {Roelens}, {Romero-G{\'o}mez}, {Rowell},
  {Royer}, {Ruiz-Dern}, {Sadowski}, {Sagrist{\`a} Sell{\'e}s}, {Sahlmann},
  {Salgado}, {Salguero}, {Sanna}, {Santana-Ros}, {Sarasso}, {Savietto},
  {Schultheis}, {Sciacca}, {Segol}, {Segovia}, {S{\'e}gransan}, {Shih},
  {Siltala}, {Silva}, {Smart}, {Smith}, {Solano}, {Solitro}, {Sordo}, {Soria
  Nieto}, {Souchay}, {Spagna}, {Spoto}, {Stampa}, {Steele},
  {Steidelm{\"u}ller}, {Stephenson}, {Stoev}, {Suess}, {Surdej}, {Szabados},
  {Szegedi-Elek}, {Tapiador}, {Taris}, {Tauran}, {Taylor}, {Teixeira},
  {Terrett}, {Teyssand ier}, {Thuillot}, {Titarenko}, {Torra Clotet}, {Turon},
  {Ulla}, {Utrilla}, {Uzzi}, {Vaillant}, {Valentini}, {Valette}, {van Elteren},
  {Van Hemelryck}, {van Leeuwen}, {Vaschetto}, {Vecchiato}, {Veljanoski},
  {Viala}, {Vicente}, {Vogt}, {von Essen}, {Voss}, {Votruba}, {Voutsinas},
  {Walmsley}, {Weiler}, {Wertz}, {Wevers}, {Wyrzykowski}, {Yoldas},
  {{\v{Z}}erjal}, {Ziaeepour}, {Zorec}, {Zschocke}, {Zucker}, {Zurbach}, \&
  {Zwitter}}]{GaiaCollaboration2018}
{Gaia Collaboration}, {Brown}, A.~G.~A., {Vallenari}, A., {et~al.} 2018, \aap,
  616, A1

\bibitem[{{Garufi} {et~al.}(2019){Garufi}, {Podio}, {Bacciotti}, {Antoniucci},
  {Boccaletti}, {Codella}, {Dougados}, {M{\'e}nard}, {Mesa}, {Meyer}, {Nisini},
  {Schmid}, {Stolker}, {Baudino}, {Biller}, {Bonavita}, {Bonnefoy},
  {Cantalloube}, {Chauvin}, {Cheetham}, {Desidera}, {D'Orazi}, {Feldt},
  {Galicher}, {Grandjean}, {Gratton}, {Hagelberg}, {Lagrange}, {Langlois},
  {Lannier}, {Lazzoni}, {Maire}, {Perrot}, {Rickman}, {Schmidt}, {Vigan},
  {Zurlo}, {Delboulb{\'e}}, {Le Mignant}, {Fantinel}, {M{\"o}ller-Nilsson},
  {Weber}, \& {Sauvage}}]{Garufi2019}
{Garufi}, A., {Podio}, L., {Bacciotti}, F., {et~al.} 2019, \aap, 628, A68

\bibitem[{{Giannini} {et~al.}(2019){Giannini}, {Nisini}, {Antoniucci},
  {Biazzo}, {Alcal{\'a}}, {Bacciotti}, {Fedele}, {Frasca}, {Harutyunyan},
  {Munari}, {Rigliaco}, \& {Vitali}}]{Giannini2019}
{Giannini}, T., {Nisini}, B., {Antoniucci}, S., {et~al.} 2019, \aap, 631, A44

\bibitem[{{Gressel} {et~al.}(2020){Gressel}, {Ramsey}, {Brinch}, {Nelson},
  {Turner}, \& {Bruderer}}]{Gressel2020}
{Gressel}, O., {Ramsey}, J.~P., {Brinch}, C., {et~al.} 2020, arXiv e-prints,
  arXiv:2005.03431

\bibitem[{{Guilloteau} {et~al.}(2016){Guilloteau}, {Reboussin}, {Dutrey},
  {Chapillon}, {Wakelam}, {Pi{\'e}tu}, {Di Folco}, {Semenov}, \&
  {Henning}}]{Guilloteau2016}
{Guilloteau}, S., {Reboussin}, L., {Dutrey}, A., {et~al.} 2016, \aap, 592, A124

\bibitem[{{Gullbring} {et~al.}(1998){Gullbring}, {Hartmann}, {Brice{\~n}o}, \&
  {Calvet}}]{Gullbring1998}
{Gullbring}, E., {Hartmann}, L., {Brice{\~n}o}, C., \& {Calvet}, N. 1998, \apj,
  492, 323

\bibitem[{{Hartigan} {et~al.}(2004){Hartigan}, {Edwards}, \&
  {Pierson}}]{Hartigan2004}
{Hartigan}, P., {Edwards}, S., \& {Pierson}, R. 2004, \apj, 609, 261

\bibitem[{{Hartigan} {et~al.}(2007){Hartigan}, {Frank}, {Varni{\'e}re}, \&
  {Blackman}}]{Hartigan2007}
{Hartigan}, P., {Frank}, A., {Varni{\'e}re}, P., \& {Blackman}, E.~G. 2007,
  \apj, 661, 910

\bibitem[{{Hartigan} \& {Hillenbrand}(2009)}]{Hartigan2009}
{Hartigan}, P. \& {Hillenbrand}, L. 2009, \apj, 705, 1388

\bibitem[{{Herczeg} \& {Hillenbrand}(2014)}]{Herczeg2014}
{Herczeg}, G.~J. \& {Hillenbrand}, L.~A. 2014, \apj, 786, 97

\bibitem[{{Hirota} {et~al.}(2017){Hirota}, {Machida}, {Matsushita}, {Motogi},
  {Matsumoto}, {Kim}, {Burns}, \& {Honma}}]{Hirota2017}
{Hirota}, T., {Machida}, M.~N., {Matsushita}, Y., {et~al.} 2017, Nature
  Astronomy, 1, 0146

\bibitem[{{Hirth} {et~al.}(1997){Hirth}, {Mundt}, \& {Solf}}]{Hirth1997}
{Hirth}, G.~A., {Mundt}, R., \& {Solf}, J. 1997, \aaps, 126, 437

\bibitem[{{Hirth} {et~al.}(1994){Hirth}, {Mundt}, {Solf}, \& {Ray}}]{Hirth1994}
{Hirth}, G.~A., {Mundt}, R., {Solf}, J., \& {Ray}, T.~P. 1994, \apjl, 427, L99

\bibitem[{{Howard} {et~al.}(2013){Howard}, {Sandell}, {Vacca}, {Duch{\^e}ne},
  {Mathews}, {Augereau}, {Barrado}, {Dent}, {Eiroa}, {Grady}, {Kamp}, {Meeus},
  {M{\'e}nard}, {Pinte}, {Podio}, {Riviere-Marichalar}, {Roberge}, {Thi},
  {Vicente}, \& {Williams}}]{Howard2013}
{Howard}, C.~D., {Sandell}, G., {Vacca}, W.~D., {et~al.} 2013, \apj, 776, 21

\bibitem[{{Huang} {et~al.}(2017){Huang}, {{\"O}berg}, {Qi}, {Aikawa},
  {Andrews}, {Furuya}, {Guzm{\'a}n}, {Loomis}, {van Dishoeck}, \&
  {Wilner}}]{Huang2017}
{Huang}, J., {{\"O}berg}, K.~I., {Qi}, C., {et~al.} 2017, \apj, 835, 231

\bibitem[{{Itoh} {et~al.}(2008){Itoh}, {Hayashi}, {Tamura}, {Oasa}, {Hioki},
  {Fukagawa}, \& {Kudo}}]{Itoh2008}
{Itoh}, Y., {Hayashi}, M., {Tamura}, M., {et~al.} 2008, \pasj, 60, 223

\bibitem[{{Lai}(2003)}]{Lai2003}
{Lai}, D. 2003, \apjl, 591, L119

\bibitem[{{Lavalley-Fouquet} {et~al.}(2000){Lavalley-Fouquet}, {Cabrit}, \&
  {Dougados}}]{LavalleyFouquet2000}
{Lavalley-Fouquet}, C., {Cabrit}, S., \& {Dougados}, C. 2000, \aap, 356, L41

\bibitem[{{Lee} {et~al.}(2017){Lee}, {Ho}, {Li}, {Hirano}, {Zhang}, \&
  {Shang}}]{Lee2017}
{Lee}, C.-F., {Ho}, P. T.~P., {Li}, Z.-Y., {et~al.} 2017, Nature Astronomy, 1,
  0152

\bibitem[{{Lii} {et~al.}(2014){Lii}, {Romanova}, {Ustyugova}, {Koldoba}, \&
  {Lovelace}}]{Lii2014}
{Lii}, P.~S., {Romanova}, M.~M., {Ustyugova}, G.~V., {Koldoba}, A.~V., \&
  {Lovelace}, R. V.~E. 2014, \mnras, 441, 86

\bibitem[{{Long} {et~al.}(2019){Long}, {Herczeg}, {Harsono}, {Pinilla},
  {Tazzari}, {Manara}, {Pascucci}, {Cabrit}, {Nisini}, {Johnstone}, {Edwards},
  {Salyk}, {Menard}, {Lodato}, {Boehler}, {Mace}, {Liu}, {Mulders}, {Hendler},
  {Ragusa}, {Fischer}, {Banzatti}, {Rigliaco}, {van de Plas}, {Dipierro},
  {Gully-Santiago}, \& {Lopez-Valdivia}}]{Long2019}
{Long}, F., {Herczeg}, G.~J., {Harsono}, D., {et~al.} 2019, \apj, 882, 49

\bibitem[{{Lovelace} {et~al.}(2010){Lovelace}, {Romanova}, {Ustyugova}, \&
  {Koldoba}}]{Lovelace2010}
{Lovelace}, R.~V.~E., {Romanova}, M.~M., {Ustyugova}, G.~V., \& {Koldoba},
  A.~V. 2010, \mnras, 408, 2083

\bibitem[{{Magakian}(2003)}]{Magakian2003}
{Magakian}, T.~Y. 2003, \aap, 399, 141

\bibitem[{{Marconi} {et~al.}(2003){Marconi}, {Axon}, {Capetti}, {Maciejewski},
  {Atkinson}, {Batcheldor}, {Binney}, {Carollo}, {Dressel}, {Ford}, {Gerssen},
  {Hughes}, {Macchetto}, {Merrifield}, {Scarlata}, {Sparks}, {Stiavelli},
  {Tsvetanov}, \& {van der Marel}}]{Marconi2003}
{Marconi}, A., {Axon}, D.~J., {Capetti}, A., {et~al.} 2003, \apj, 586, 868

\bibitem[{{Masciadri} \& {Raga}(2002)}]{Masciadri2002}
{Masciadri}, E. \& {Raga}, A.~C. 2002, \apj, 568, 733

\bibitem[{{Matt} \& {Pudritz}(2005)}]{Matt2005}
{Matt}, S. \& {Pudritz}, R.~E. 2005, \apjl, 632, L135

\bibitem[{{Maurri} {et~al.}(2014){Maurri}, {Bacciotti}, {Podio},
  {Eisl{\"o}ffel}, {Ray}, {Mundt}, {Locatelli}, \& {Coffey}}]{Maurri2014}
{Maurri}, L., {Bacciotti}, F., {Podio}, L., {et~al.} 2014, \aap, 565, A110

\bibitem[{{McGroarty} {et~al.}(2004){McGroarty}, {Ray}, \&
  {Bally}}]{McGroarty2004}
{McGroarty}, F., {Ray}, T.~P., \& {Bally}, J. 2004, \aap, 415, 189

\bibitem[{{Melnikov} {et~al.}(2009){Melnikov}, {Eisl{\"o}ffel}, {Bacciotti},
  {Woitas}, \& {Ray}}]{Melnikov2009}
{Melnikov}, S.~Y., {Eisl{\"o}ffel}, J., {Bacciotti}, F., {Woitas}, J., \&
  {Ray}, T.~P. 2009, \aap, 506, 763

\bibitem[{{Nealon} {et~al.}(2018){Nealon}, {Dipierro}, {Alexander}, {Martin},
  \& {Nixon}}]{Nealson2018}
{Nealon}, R., {Dipierro}, G., {Alexander}, R., {Martin}, R.~G., \& {Nixon}, C.
  2018, \mnras, 481, 20

\bibitem[{{Nisini} {et~al.}(2005){Nisini}, {Bacciotti}, {Giannini}, {Massi},
  {Eisl{\"o}ffel}, {Podio}, \& {Ray}}]{Nisini2005}
{Nisini}, B., {Bacciotti}, F., {Giannini}, T., {et~al.} 2005, \aap, 441, 159

\bibitem[{{Panoglou} {et~al.}(2012){Panoglou}, {Cabrit}, {Pineau Des
  For{\^e}ts}, {Garcia}, {Ferreira}, \& {Casse}}]{Panoglou2012}
{Panoglou}, D., {Cabrit}, S., {Pineau Des For{\^e}ts}, G., {et~al.} 2012, \aap,
  538, A2

\bibitem[{{Papaloizou} \& {Terquem}(1995)}]{Papaloizou1995}
{Papaloizou}, J. C.~B. \& {Terquem}, C. 1995, \mnras, 274, 987

\bibitem[{{Pesenti} {et~al.}(2004){Pesenti}, {Dougados}, {Cabrit}, {Ferreira},
  {Casse}, {Garcia}, \& {O'Brien}}]{Pesenti2004}
{Pesenti}, N., {Dougados}, C., {Cabrit}, S., {et~al.} 2004, \aap, 416, L9

\bibitem[{{Podio} {et~al.}(2011){Podio}, {Eisl{\"o}ffel}, {Melnikov}, {Hodapp},
  \& {Bacciotti}}]{Podio2011}
{Podio}, L., {Eisl{\"o}ffel}, J., {Melnikov}, S., {Hodapp}, K.~W., \&
  {Bacciotti}, F. 2011, \aap, 527, A13

\bibitem[{{Pradhan} \& {Zhang}(1993)}]{Pradhan1993}
{Pradhan}, A.~K. \& {Zhang}, H.~L. 1993, \apjl, 409, L77

\bibitem[{{Pudritz} {et~al.}(2007){Pudritz}, {Ouyed}, {Fendt}, \&
  {Brandenburg}}]{Pudritz2007}
{Pudritz}, R.~E., {Ouyed}, R., {Fendt}, C., \& {Brandenburg}, A. 2007, in
  Protostars and Planets V, ed. B.~{Reipurth}, D.~{Jewitt}, \& K.~{Keil}, 277

\bibitem[{{Ray} {et~al.}(2007){Ray}, {Dougados}, {Bacciotti}, {Eisl{\"o}ffel},
  \& {Chrysostomou}}]{Ray2007}
{Ray}, T., {Dougados}, C., {Bacciotti}, F., {Eisl{\"o}ffel}, J., \&
  {Chrysostomou}, A. 2007, in Protostars and Planets V, ed. B.~{Reipurth},
  D.~{Jewitt}, \& K.~{Keil}, 231

\bibitem[{{Riols} {et~al.}(2020){Riols}, {Lesur}, \& {Menard}}]{Riols2020}
{Riols}, A., {Lesur}, G., \& {Menard}, F. 2020, arXiv e-prints,
  arXiv:2006.01194

\bibitem[{{Rodriguez} {et~al.}(2018){Rodriguez}, {Loomis}, {Cabrit}, {Haworth},
  {Facchini}, {Dougados}, {Booth}, {Jensen}, {Clarke}, {Stassun}, {Dent}, \&
  {Pety}}]{Rodriguez2018}
{Rodriguez}, J.~E., {Loomis}, R., {Cabrit}, S., {et~al.} 2018, \apj, 859, 150

\bibitem[{{Romanova} {et~al.}(2009){Romanova}, {Ustyugova}, {Koldoba}, \&
  {Lovelace}}]{Romanova2009}
{Romanova}, M.~M., {Ustyugova}, G.~V., {Koldoba}, A.~V., \& {Lovelace},
  R.~V.~E. 2009, \mnras, 399, 1802

\bibitem[{{Shakura} \& {Sunyaev}(1973)}]{Shakura1973}
{Shakura}, N.~I. \& {Sunyaev}, R.~A. 1973, \aap, 500, 33

\bibitem[{{Sheikhnezami} \& {Fendt}(2015)}]{SheikhnezamiFendt2015}
{Sheikhnezami}, S. \& {Fendt}, C. 2015, \apj, 814, 113

\bibitem[{{Sheikhnezami} \& {Fendt}(2018)}]{SheikhnezamiFendt2018}
{Sheikhnezami}, S. \& {Fendt}, C. 2018, \apj, 861, 11

\bibitem[{{Shu} {et~al.}(2000){Shu}, {Najita}, {Shang}, \& {Li}}]{Shu2000}
{Shu}, F.~H., {Najita}, J.~R., {Shang}, H., \& {Li}, Z.~Y. 2000, in Protostars
  and Planets IV, ed. V.~{Mannings}, A.~P. {Boss}, \& S.~S. {Russell}, 789--814

\bibitem[{{Simon} {et~al.}(2016){Simon}, {Pascucci}, {Edwards}, {Feng},
  {Gorti}, {Hollenbach}, {Rigliaco}, \& {Keane}}]{Simon2016}
{Simon}, M.~N., {Pascucci}, I., {Edwards}, S., {et~al.} 2016, \apj, 831, 169

\bibitem[{{Soker}(2005)}]{Soker2005}
{Soker}, N. 2005, \aap, 435, 125

\bibitem[{{Staff} {et~al.}(2012){Staff}, {Jaikumar}, {Chan}, \&
  {Ouyed}}]{Staff2012}
{Staff}, J.~E., {Jaikumar}, P., {Chan}, V., \& {Ouyed}, R. 2012, \apj, 751, 24

\bibitem[{{Tabone} {et~al.}(2020){Tabone}, {Cabrit}, {Pineau des For{\^e}ts},
  {Ferreira}, {Gusdorf}, {Podio}, {Bianchi}, {Chapillon}, {Codella}, \&
  {Gueth}}]{Tabone2020}
{Tabone}, B., {Cabrit}, S., {Pineau des For{\^e}ts}, G., {et~al.} 2020, arXiv
  e-prints, arXiv:2004.08804

\bibitem[{{Terquem} {et~al.}(1999){Terquem}, {Eisl{\"o}ffel}, {Papaloizou}, \&
  {Nelson}}]{Terquem1999}
{Terquem}, C., {Eisl{\"o}ffel}, J., {Papaloizou}, J.~C.~B., \& {Nelson}, R.~P.
  1999, \apjl, 512, L131

\bibitem[{{Terquem} {et~al.}(1998){Terquem}, {Papaloizou}, {Nelson}, \&
  {Lin}}]{Terquem1998}
{Terquem}, C., {Papaloizou}, J.~C.~B., {Nelson}, R.~P., \& {Lin}, D.~N.~C.
  1998, \apj, 502, 788

\bibitem[{{Turner} {et~al.}(2014){Turner}, {Fromang}, {Gammie}, {Klahr},
  {Lesur}, {Wardle}, \& {Bai}}]{Turner2014}
{Turner}, N.~J., {Fromang}, S., {Gammie}, C., {et~al.} 2014, in Protostars and
  Planets VI, ed. H.~{Beuther}, R.~S. {Klessen}, C.~P. {Dullemond}, \&
  T.~{Henning}, 411

\bibitem[{{Uchida} \& {Shibata}(1985)}]{UchidaShibata1985}
{Uchida}, Y. \& {Shibata}, K. 1985, \pasj, 37, 515

\bibitem[{{Whelan} \& {Garcia}(2008)}]{Whelan2008}
{Whelan}, E. \& {Garcia}, P. 2008, {Spectro-astrometry: The Method, its
  Limitations, and Applications}, Vol. 742 (Springer), 123

\bibitem[{{White} {et~al.}(2014){White}, {McGregor}, {Bicknell}, {Salmeron}, \&
  {Beck}}]{White2014}
{White}, M.~C., {McGregor}, P.~J., {Bicknell}, G.~V., {Salmeron}, R., \&
  {Beck}, T.~L. 2014, \mnras, 441, 1681

\bibitem[{{Winter} {et~al.}(2018){Winter}, {Booth}, \& {Clarke}}]{Winter2018}
{Winter}, A.~J., {Booth}, R.~A., \& {Clarke}, C.~J. 2018, \mnras, 479, 5522

\bibitem[{{Woitas} {et~al.}(2005){Woitas}, {Bacciotti}, {Ray}, {Marconi},
  {Coffey}, \& {Eisl{\"o}ffel}}]{Woitas2005}
{Woitas}, J., {Bacciotti}, F., {Ray}, T.~P., {et~al.} 2005, \aap, 432, 149

\bibitem[{{Woitas} {et~al.}(2002){Woitas}, {Ray}, {Bacciotti}, {Davis}, \&
  {Eisl{\"o}ffel}}]{Woitas2002}
{Woitas}, J., {Ray}, T.~P., {Bacciotti}, F., {Davis}, C.~J., \&
  {Eisl{\"o}ffel}, J. 2002, \apj, 580, 336

\bibitem[{{Xiang-Gruess} \& {Papaloizou}(2013)}]{Xiang-Gruess2013}
{Xiang-Gruess}, M. \& {Papaloizou}, J.~C.~B. 2013, \mnras, 431, 1320

\bibitem[{{Zanni} \& {Ferreira}(2013)}]{Zanni2013}
{Zanni}, C. \& {Ferreira}, J. 2013, \aap, 550, A99

\bibitem[{{Zhu}(2019)}]{Zhu2019}
{Zhu}, Z. 2019, \mnras, 483, 4221

\end{thebibliography}


\appendix
\section{Uneven slit illumination}

For accurate rotation measurements, our derived \ion{Fe}{II} velocity centroid maps need to be corrected for spurious velocity shifts due to uneven-slit illumination. The slicing mirrors in NIFS/GEMINI behave like a long-slit spectrograph, for which off-axis light suffers from a shift in wavelength on the detector with respect to on-axis light. The magnitude of this effect depends on both the centering and gradient of light distribution within the “slitlet”. In our observations, the total light distribution is dominated by the PSF of the stellar continuum, which has a FWHM on the order of the “slitlet” width (0.1$^{\prime\prime}$). Therefore, one might expect this effect to be significant over the spatial extent of the PSF wings which, due to imperfect adaptive optics (AO) correction, cover a significant fraction of the NIFS field of view.

We followed the method outlined in appendix~A of \citet{Agra-Amboage2014} to model the uneven slit illumination effect. We take, as the incoming light distribution, an image in a wavelength band which includes the relevant emission line plus the nearby continuum. This is obtained by spectrally integrating the datacube in a narrow wavelength domain centered on the line of interest, here [\ion{Fe}{II}]$\lambda$1.64 $\mu$m, and over-sampled by a factor 5. In each spaxel, we then estimate the location of the brightness centroid and its displacement with respect to the slitlet center. The velocity shift induced by the uneven slit illumination effect on a given spaxel is then given by the following formula, adapted from \citet{Marconi2003}: 

\begin{equation*}
    \Delta v  (x_0,y_0) = \frac{\delta v}{\delta x/2.} \times \frac{\iint I_{mod}(u,v) \times (u-x_0) \,du\,dv}{\iint I_{mod}(u,v) \,du\,dv}
\end{equation*}
 
 where $\delta v$ is the spectral pixel sampling, $\delta x$ the width of the slit, and $I_{\rm mod}$ is the model infalling light intensity integrated over the current spaxel of detector coordinates (x$_0$,y$_0$).

\begin{figure*}
    \centering
    \includegraphics[width=0.4\textwidth]{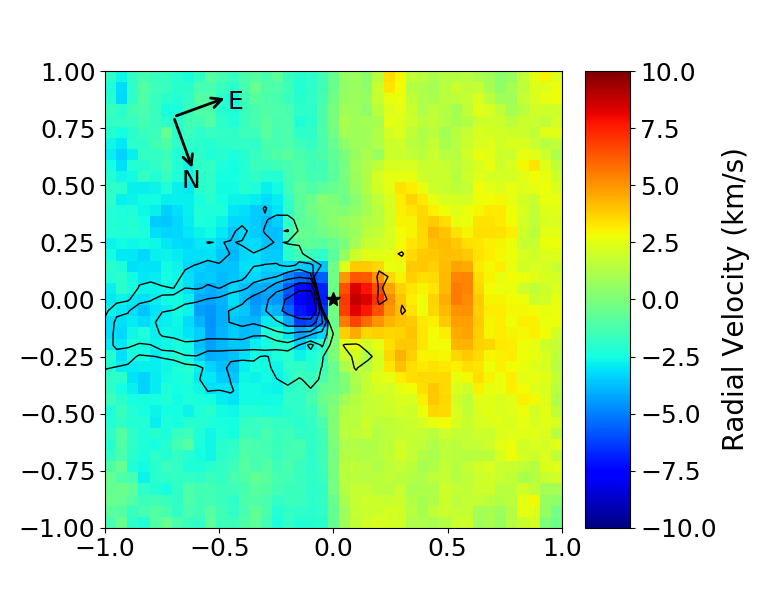} 
    \includegraphics[width=0.4\textwidth]{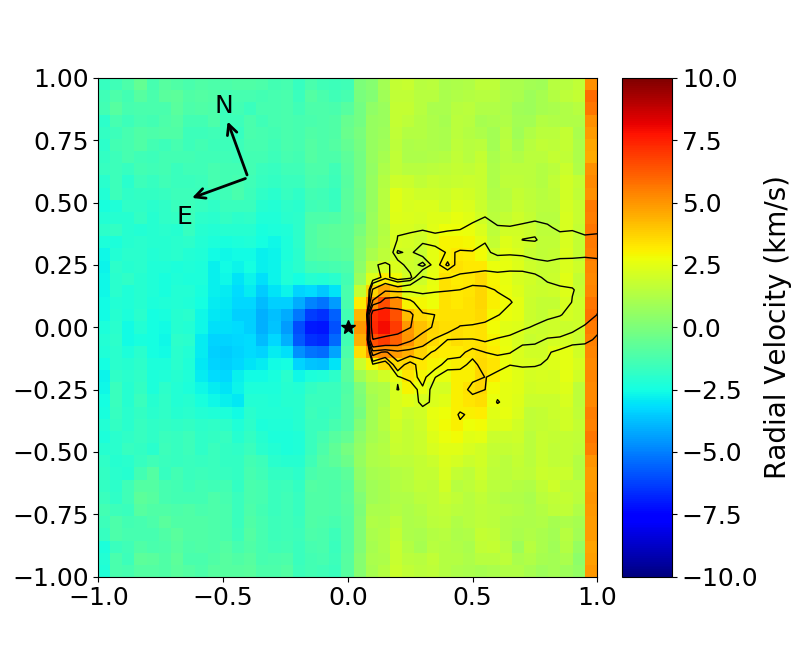} 
    \includegraphics[width=0.4\textwidth,trim=3cm 1cm 3cm 1cm]{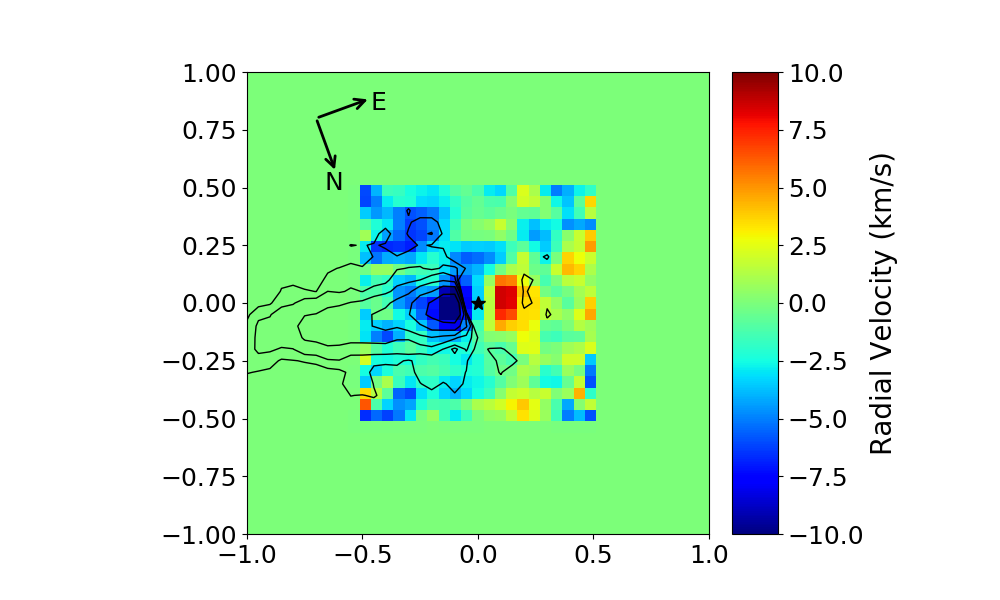} 
    \includegraphics[width=0.4\textwidth,trim=3cm 1cm 3cm 1cm]{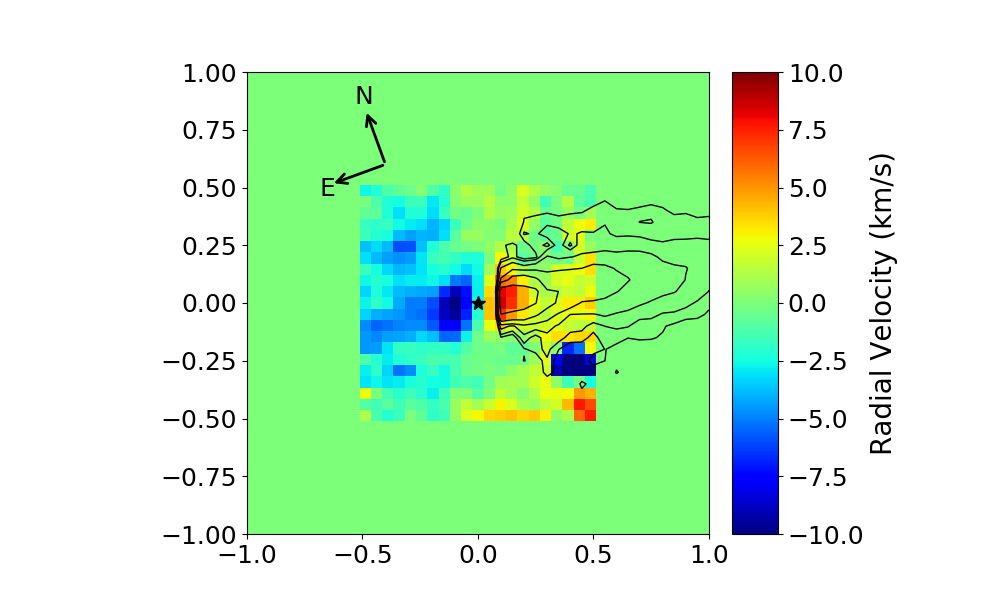} 
    \caption{Maps of velocity corrections required as a result of uneven slit illumination, for both PA 160 \degr (left) and PA 340 \degr (right). Top panels are computed by modelling the uneven slit illumination effect in each spaxel. Bottom panels are measured velocity centroid shifts relative to the central spectrum for the photospheric line near 16755$\angstrom$, as a double-check on the model output. Black contours show the position of the approaching jet.}
    \label{fig:uneven_slit} 
\end{figure*}

Figure \ref{fig:uneven_slit} shows the resulting
2D maps of the velocity corrections around the [\ion{Fe}{II}]$\lambda$1.64$\mu$m line for the datacubes corresponding to the two instrument position angles used. These velocity corrections reach 7-10 km s$^{-1}$ close to the stellar peak position, and rapidly decrease to a few km s$^{-1}$ at larger distances. As expected the velocity corrections have a stronger variation along the $x$ (dispersion) direction. 

We also determined empirically the velocity shifts induced by the uneven slit effect by looking at a photospheric absorption line near 16755~$\angstrom$, located in the vicinity of the [\ion{Fe}{II}] 1.64$\mu$m line. Figure ~\ref{fig:uneven_slit} shows the centroid velocity shifts  relative to the central spectrum ($v_{\rm cen}(x,y) - v_{\rm cen}(x_0,y_0)$), derived by Gaussian fitting of the line profile. Although these maps are more noisy, they show very good agreement with the computed model velocity correction maps above, Figure \ref{fig:uneven_slit}.  

\end{document}